\documentclass[useAMS,usegraphicx,usenatbib]{mn2e}

\newcommand{\Msun}{\ensuremath{\,{\rm M}_\odot}}           
\newcommand{\Teff}{\ensuremath{T_{\rm eff}}}               
\newcommand{\logg}{\ensuremath{\log g}}                    
\newcommand{\kms}{\,km\,s$^{-1}$}                          
\newcommand{\Apx}{\,\AA\,px$^{-1}$}                        
\newcommand{\ion}[2]{{#1}\,{\sc {\small{#2}}}}             
\newcommand{\mc}[1]{\multicolumn{2}{c}{#1}}
\newcommand{\cd}{\ensuremath{\,{\rm cycle\,d}^{-1}}}              

\title[VLT spectroscopy of CVs discovered by the SDSS]
      {VLT/FORS spectroscopy of faint cataclysmic variables discovered by the
      Sloan Digital Sky Survey}

\author[Southworth et al.]
       {John Southworth$^1$\thanks{E-mail: j.k.taylor@warwick.ac.uk (JS),
        \newline Boris.Gaensicke@warwick.ac.uk (BTG),
        \newline T.R.Marsh@warwick.ac.uk (TRM)},
        B.\ T.\ G\"ansicke$^1$, 
        T.\ R.\ Marsh$^1$, 
        D. de Martino$^2$,
        P. Hakala$^3$,\newauthor
        S. Littlefair$^4$,
        P. Rodr{\'i}guez-Gil$^5$, 
        P. Szkody$^6$ 
        \\
        $^1$ Department of Physics, University of Warwick, Coventry, CV4 7AL, UK          \\
        $^2$ INAF - Osservatorio di Capodimonte, Via Moiariello 16, 80131 Napoli, Italy   \\
        $^3$ Observatory, University of Helsinki, PO Box 14, Helsinki, Finland            \\
        $^4$ Department of Physics and Astronomy, University of Sheffield, S3 7RH, UK     \\
        $^5$ Instituto de Astrof\'isica de Canarias, 38200 La Laguna, Tenerife, Spain     \\
        $^6$ Astronomy Department, University of Washington, Seattle, WA98195, USA         }

\begin{document} \maketitle 

\begin{abstract}
We present medium-resolution VLT/FORS2 spectroscopy of six cataclysmic variables discovered by the Sloan Digital Sky Survey. We determine orbital periods for SDSS\,J023322.61+005059.5 ($96.08\pm0.09$\,min), SDSS\,J091127.36+084140.7 ($295.74\pm0.22$\,min), SDSS\,J103533.02+055158.3 ($82.10\pm0.09$\,min), and SDSS\,J121607.03+052013.9 (most likely $98.82\pm0.16$\,min, but the one-day aliases at 92\,min and 107\,min are also possible) using radial velocities measured from their H$\alpha$ and H$\beta$ emission lines. Three of the four orbital periods measured here are close to the observed 75--80\,min minimum period for cataclysmic variables, indicating that the properties of the population of these objects discovered by the SDSS are substantially different to those of the cataclysmic variables found by other means. Additional photometry of SDSS\,J023322.61+005059.5 reveals a periodicity of approximately 60\,min which we interpret as the spin period of the white dwarf, suggesting that this system is an intermediate polar with a low accretion rate. SDSS\,J103533.02+055158.3 has a period right at the observed minimum value, a spectrum dominated by the cool white dwarf primary star and exhibits deep eclipses, so is an excellent candidate for an accurate determination of the parameters of the system. The spectroscopic orbit of SDSS\,J121607.03+052013.9 has a velocity amplitude of only $13.8 \pm 1.6$\kms, implying that this system has an extreme mass ratio. From several physical constraints we find that this object must contain either a high-mass white dwarf or a brown-dwarf-mass secondary component or both.
\end{abstract}

\begin{keywords}
stars: novae, cataclysmic variables -- stars: binaries: close -- stars: binaries: eclipsing -- stars: white dwarfs -- stars: individual: SDSS J023322.61+005059.5 -- stars: individual: SDSS J091127.36+084140.7 -- stars: individual: SDSS J093238.21+010902.5 -- stars: individual: SDSS J101037.05+024914.9 -- stars: individual: SDSS J103533.02+055158.3 -- stars: individual: SDSS J121607.03+052013.9 -- stars: individual: SDSS J133941.11+484727.5
\end{keywords}

\section{Introduction}                                                       \label{sec:intro}

Cataclysmic variables (CVs) are interacting binary stars containing a white dwarf primary star and a low-mass secondary star in a close orbit. In the majority of systems the secondary star is unevolved, fills its Roche lobe and loses mass to the primary star via an accretion disc. Comprehensive reviews of the properties of CVs have been given by \citet{Warner95book} and \citet{Hellier01book}.

The evolution of CVs depends primarily on angular momentum loss (AML) from their orbit. CVs with long periods contain a secondary star which is only partially convective. It is thought that the presence of a magnetic field in this star causes AML by magnetic braking (\citealt{VerbuntZwaan81aa}; \citealt*{Rappaport++82apj}) until the orbital period decreases to about three hours. At this point the secondary star becomes fully convective and magnetic braking ceases. This change in the state of the star causes it to relax to its thermal equilibrium radius and so shrink inside its Roche lobe. Mass transfer ceases and weaker AML is now dominated by gravitational radiation \citep{Paczynski67aca}. Once the period reaches about two hours, the shrinking Roche lobe re-establishes contact with the secondary star, restarting the mass transfer. The period continues to decrease to an observed minimum of approximately 80 minutes, at which point the secondary star becomes a degenerate brown-dwarf like object. As the mass donor is now degenerate, further mass transfer causes the period to increase \citep{Patterson98pasp}.

Systems beyond this ''period bounce'' are theoretically predicted to represent up to 70\% of the steady-state population of CVs; population synthesis models also predict that the vast majority (99\%) of CVs have evolved to periods shorter than the 2--3 hour gap \citep{Dekool92aa,DekoolRitter93aa,Kolb93aa,Politano96apj,KolbBaraffe99mn}. The models also predict that the low rate of change of period near the minimum value causes an increase in the population of CVs there, which should manifest itself as a spike in the period distribution. However, whilst several candidates exist, not one CV has yet been confirmed to have a secondary component with a mass appropriate for a period bounce CV \citep*{Littlefair++03mn,Patterson++05pasp}. In addition, the observed period distribution of CVs does not have a spike at the minimum period and it does not contain far more short-period systems than those with periods beyond the well-known 2--3 hour gap \citep[e.g.][]{Downes+01pasp,RitterKolb03aa}. Furthermore, the theoretical minimum period is generally found to be around 65 minutes, which is substantially shorter than the observed value. Many different AML mechanisms have been put forward as possible solutions to these problems (e.g.\ \citealt*{Andronov++03apj}; \citealt{TaamSpruit01apj,Schenker+02mn}), without complete success. The AML rates given by different prescriptions can differ by several orders of magnitude \citep{SchreiberGansicke03aa}.

Due to these difficulties, detailed observational population studies are of great importance to provide the constraints necessary to further develop the theory of compact binary evolution. The currently known population of CVs is afflicted by several strong biases which makes the characterisation of the true population -- and a comparison with models -- extremely difficult.  The Sloan Digital Sky Survey (SDSS\footnote{\tt http://www.sdss.org/}; \citealt{York+00aj}) is a large project to obtain five-colour imaging and multi-object spectroscopy of a large area at high galactic latitudes in the Northern sky, with the main science goal being the measurement of redshifts for an extensive sample of galaxies and quasars. Objects for spectroscopic follow-up are chosen, on the basis of their photometric colours, to be blue and/or distant from the locus of the stellar main sequence. Approximately 200 CVs, with a wide range of photometric colours, have been detected in the SDSS spectroscopic database \citep{Szkody+02aj, Szkody+03aj, Szkody+04aj, Szkody+05aj, Szkody+06aj}, and many more are expected to be found from future SDSS observations. The limiting magnitude of this population is $g \sim 20$, much fainter than previous surveys (e.g.\ \citealt*{Green++86apjs}), so the SDSS CV population is both more homogeneous and on average fainter than the currently known sample of CVs. It is therefore expected to contain more short-period systems, and may include a number of intrinsically faint CVs which have evolved beyond the period bounce.

We are conducting a research program to measure the orbital periods of SDSS CVs (\citealt{Gansicke05aspc, Gansicke+06mn}; Dillon et al., 2006, in preparation) in order to assess whether the study of a less biased sample of CVs can reconcile observations with theories of the evolution of interacting binary stars. In this work we present phase-resolved spectroscopy of six faint SDSS CVs (Table~\ref{tab:iddata}) and measure orbital periods for SDSS\,J023322.6+005059.5, SDSS\,J091127.36+084140.7, SDSS\,J103533.02+055158.3 and SDSS\,J121607.03+052013.9. The spectra of SDSS\,J103533.02+055158.3 show that it is an eclipsing system. In addition we observed SDSS\,J101037.05+024914.9, finding that it was several magnitudes fainter than in the SDSS observations, and SDSS\,J093238.2+010902.5, but did not detect a coherent radial velocity motion. Our selection policy was to pick faint CVs which were discovered by the SDSS and were observable from the VLT. We did not attempt to carefully select a representative sample of objects because the goal of our research project is to (eventually) measure the orbital periods of {\it all} of the SDSS CVs with unknown periods. In this work we shall abbreviate the names of the targets to SDSS\,J0233, SDSS\,J0911, SDSS\,J0935, SDSS\,J1010, SDSS\,J1035 and SDSS\,J1216.

\begin{table*} \begin{center}
\caption{\label{tab:iddata} Apparent magnitudes of our targets in the SDSS $ugriz$
passbands. $r_{\rm spec}$ is an apparent magnitude calculated by the SDSS by convolving
their flux-calibrated spectra with the $r$ passband function, and is obtained at a
different epoch to the $ugriz$ magnitudes measured from the CCD imaging observations.}
\begin{tabular}{lllccccccc} \hline
SDSS name                & Short name & Reference     & $u$   & $g$   & $r$   & $i$   & $z$   & $r_{\rm spec}$ \\
\hline
SDSS J023322.61+005059.5 & SDSS\,J0233 & \citet{Szkody+02aj} & 19.76 & 19.86 & 19.73 & 19.93 & 19.82 & 19.92 \\
SDSS J091127.36+084140.7 & SDSS\,J0911 & \citet{Szkody+05aj} & 19.63 & 19.70 & 19.16 & 18.66 & 18.28 & 19.51 \\
SDSS J093238.21+010902.5 & SDSS\,J0932 & \citet{Szkody+03aj} & 19.70 & 20.26 & 19.53 & 19.10 & 19.12 & 19.71 \\
SDSS J101037.05+024914.9 & SDSS\,J1010 & \citet{Szkody+03aj} & 20.39 & 20.75 & 20.35 & 20.71 & 20.46 & 20.63 \\
SDSS J103533.02+055158.3 & SDSS\,J1035 & \citet{Szkody+06aj} & 19.14 & 18.79 & 18.81 & 18.97 & 19.12 & 18.99 \\
SDSS J121607.03+052013.9 & SDSS\,J1216 & \citet{Szkody+04aj} & 20.32 & 20.13 & 19.98 & 20.26 & 20.27 & 20.10 \\
\hline \end{tabular} \end{center} \end{table*}


\section{Observations and data reduction}\label{sec:obs}

\begin{table*} \begin{center}
\caption{\label{tab:obslog} Log of the observations presented in this work. The acquisition magnitudes were measured from the VLT/FORS2 acquisition images and are discussed in Section~\ref{sec:obs:vltphot}.}
\begin{tabular}{lcccccrrr} \hline
Target & Date & Start time & End time & Telescope and & Optical  & Number of   & Exposure & Acquisition \\
       & (UT) &  (UT)      &  (UT)    &  instrument   & element  &observations & time (s) & magnitude   \\
\hline
SDSS\,J0233 & 2004 08 25 & 03 22 & 05 50 & LT RATCAM & SDSS $g$ filter & 101 &  60 &      \\
SDSS\,J0233 & 2004 08 28 & 03 41 & 05 55 & LT RATCAM & SDSS $g$ filter &  92 &  60 &      \\
SDSS\,J0233 & 2005 01 02 & 20 53 & 23 55 & INT WFC   & SDSS $g$ filter &  80 & 60--120 &  \\
SDSS\,J0233 & 2005 01 03 & 20 04 & 23 41 & INT WFC   & SDSS $g$ filter & 113 &  70 &      \\
SDSS\,J0233 & 2005 01 04 & 19 54 & 22 54 & INT WFC   & SDSS $g$ filter & 106 & 50--80 &   \\
SDSS\,J0233 & 2005 01 04 & 20 38 & 22 45 & WHT ISIS  & R600B R316R &   9 & 900 &      \\
SDSS\,J0233 & 2005 01 06 & 19 59 & 23 48 & WHT ISIS  & R600B R316R &  10 & 900 &      \\
SDSS\,J0233 & 2005 01 07 & 00 08 & 23 53 & WHT ISIS  & R600B R316R &  15 & 900 &      \\
SDSS\,J0233 & 2006 01 26 & 00 36 & 03 06 & VLT FORS2 & 1400V grism &  18 & 480 & 19.6 \\
SDSS\,J0233 & 2006 01 27 & 00 29 & 01 49 & VLT FORS2 & 1400V grism &  10 & 480 & 19.9 \\
SDSS\,J0233 & 2006 01 28 & 02 18 & 03 02 & VLT FORS2 & 1400V grism &  10 & 480 & 19.9 \\
\hline
SDSS\,J0911 & 2006 01 26 & 04 22 & 07 04 & VLT FORS2 & 1200R grism &  19 & 480 & 18.4 \\
SDSS\,J0911 & 2006 01 27 & 03 40 & 05 18 & VLT FORS2 & 1200R grism &  12 & 480 & 18.4 \\
SDSS\,J0911 & 2006 01 27 & 06 54 & 08 22 & VLT FORS2 & 1200R grism &  11 & 480 & 18.4 \\
\hline
SDSS\,J0932 & 2006 01 27 & 02 09 & 03 21 & VLT FORS2 & 1200R grism &   9 & 480 & 19.9 \\
\hline
SDSS\,J1010 & 2006 01 26 & 03 26 & 04 02 & VLT FORS2 & 1200R grism &   5 & 480 & 22.3 \\
SDSS\,J1010 & 2006 01 26 & 07 17 & 07 18 & VLT FORS2 & $B$ filter  &   1 &  60 & 22.0 \\
SDSS\,J1010 & 2006 01 26 & 07 19 & 07 20 & VLT FORS2 & $V$ filter  &   1 &  60 & 21.9 \\
\hline
SDSS\,J1035 & 2006 01 27 & 05 40 & 07 08 & VLT FORS2 & 1200R grism &  22 & 200 & 18.8 \\
SDSS\,J1035 & 2006 01 28 & 03 22 & 04 40 & VLT FORS2 & 1200R grism &  20 & 200 & 18.7 \\
SDSS\,J1035 & 2006 01 28 & 08 39 & 09 40 & VLT FORS2 & 1200R grism &  16 & 200 & 18.7 \\
\hline
SDSS\,J1216 & 2006 01 26 & 07 32 & 09 35 & VLT FORS2 & 1200R grism &  14 & 480 & 19.5 \\
SDSS\,J1216 & 2006 01 27 & 07 22 & 09 38 & VLT FORS2 & 1200R grism &  16 & 480 & 19.7 \\
SDSS\,J1216 & 2006 01 28 & 04 56 & 06 38 & VLT FORS2 & 1200R grism &  11 & 480 & 19.8 \\
\hline \end{tabular} \end{center} \end{table*}

\subsection{VLT spectroscopy}\label{sec:obs:vltspec}

Spectroscopic observations were carried out in 2006 January using the FORS2 spectrograph on Unit Telescope 1 of the Very Large Telescope (VLT) at ESO Paranal, Chile. The 1400V grism was used for the observations of SDSS\,J0233 and the 1200R grism for observations of the other targets. Observations using the 1400V grism cover the wavelength interval 4630 to 5930\,\AA, which includes H$\beta$, at a reciprocal dispersion of 0.64\,\Apx. Those using the 1200R grism were targeted at H$\alpha$ and cover the wavelength interval 5870 to 7370\,\AA\ at 0.73\Apx. A log of observations is given in Table~\ref{tab:obslog}. From measurements of the full widths at half maximum (FWHMs) of arc-lamp and night-sky spectral emission lines, we estimate that the resolution of our observations is 1.2\,\AA\ at H$\beta$ and 1.5\,\AA\ at H$\alpha$.

Data reduction was undertaken using optimal extraction (\citealt{Horne86pasp}) as implemented in the {\sc pamela}\footnote{{\sc pamela} and {\sc molly} were written by TRM and can be found at {\tt http://quetzel.csc.warwick.ac.uk/phsaap/software/}} code (\citealt{Marsh89pasp}), which also makes use of the {\sc starlink}\footnote{The Starlink Software Group homepage is at {\tt http://www.starlink.rl.ac.uk/}} packages {\sc figaro} and {\sc kappa}.

The wavelength calibration of the spectra was undertaken using the {\sc molly}$^2$ program and arc lamp exposures taken during daytime as part of the standard calibration routines for FORS2. However, we have found that flexure of the spectrograph can cause the position of the spectrum on the CCD to vary by up to 0.6 pixels (about 20\kms) depending on elevation. We have removed this trend by applying pixel shifts to the wavelength calibrations of the spectra to force the centroids of the night sky emission lines at 5577.338 and 6300.304\,\AA\ \citep{Osterbrock+96pasp} to these wavelengths. We have been able to check this procedure in some observations when other night sky emission lines are strong, and find that it is always accurate to within 0.1 pixels.

\subsection{VLT photometry}\label{sec:obs:vltphot}

The spectroscopic observing procedure of FORS2 included obtaining target acquisition images from which photometry can be obtained. Exposure times were generally 20\,s and the observations were unfiltered, which means that the Poisson noise is in all cases smaller than the systematic errors. We have extracted differential photometry from these images using the {\sc iraf}\footnote{{\sc iraf} is distributed by the National Optical Astronomical Observatories, which are operated by the Association of Universities for Research in Astronomy, Inc.\ under contract with the National Science Foundation.} tool {\sc imexamine}, taking special care to select comparison stars from the SDSS database with colours as close to our target CVs as possible in order to minimise colour effects for those acquisition images taken in white light. The resulting magnitudes (Table~\ref{tab:obslog}) approximate the SDSS $r$ passband, with some uncertainty due to the fact that CVs have quite different spectral energy distributions from the comparison stars.

Images in the Johnson $B$ and $V$ passbands were also obtained for SDSS\,J1010 after we found that spectroscopic observations of it had a much lower signal than expected. We used the colour transformations by R.\ Lupton on the SDSS website\footnote{Equations are given relating $ugri$ to $B$ and $V$, and are attributed to Lupton (2005) without further information about this reference. For further information see \\ {\tt http://www.sdss.org/dr4/algorithms/sdssUBVRITransform.html}} to convert the $ugri$ magnitudes of the comparison stars to the $B$ and $V$ magnitudes needed for our differential photometry.

\subsection{WHT spectroscopy}

We obtained time-resolved spectroscopy of SDSS\,J0233 at the 4\,m William Herschel Telescope (WHT) on La Palma in 2005 January, using the double-arm spectrograph ISIS with the R600B grating and 4k$\times$2k pixel EEV detector in the blue arm,  the R316R grating and 4.5k$\times$2k pixel Marconi detector in the red arm, and a 1.2'' slit width (Table~\ref{tab:obslog}). The blue arm spectra cover the wavelength range 3900--5400\,\AA\ with a resolution of 1.8\,\AA\ (corresponding to two binned pixels) and the red arm spectra cover 6100--8900\,\AA\ with a resolution of 3.6\,\AA. Regular arc lamp and flat-field exposures were obtained to adjust the wavelength calibration for the instrument flexure and to remove CCD fringing in the red spectra. The data were bias and flat-field corrected using {\sc figaro}. Optimal extraction and wavelength and flux calibration were performed in {\sc pamela} and {\sc molly}. Due to the bad atmospheric conditions (variable seeing with an average of about 5 arcseconds) the quality of the individual spectra is relatively poor. The average spectrum of SDSS\,J0233 from these observations shows strong double-peaked Balmer emission lines and relatively weak \ion{He}{I} lines.

\subsection{INT and LT photometry}

Differential $g$-band photometry of SDSS\,J0233 was obtained on the Liverpool Telescope (LT) in August 2004 and the Isaac Newton Telescope (INT) in January 2005 (Table~\ref{tab:obslog}). The INT was equipped with the Wide Field Camera (WFC), an array of four EEV 2k$\times$4k pixel CCDs, and the LT with RATCAM, which contains an EEV 2k$\times$2k pixel CCD. No binning or windowing was adopted. The data were reduced using the pipeline described by \citet{gansicke+04aa}, which performs bias and flat-field corrections within {\sc midas}\footnote{\tt http://www.eso.org/projects/esomidas/} and uses the {\sc sextractor} package \citep{BertinArnouts96aas} to perform aperture photometry for all objects in the field of view. Instrumental $g$ magnitudes were converted into apparent magnitudes using several comparison stars near the target along with their magnitudes given on the SDSS website.


\section{Data analysis}

\subsection{Radial velocity measurement}                                \label{sec:data:rv}

We measured radial velocities from the hydrogen emission lines by cross-correlation with single and double Gaussian functions, as implemented in {\sc molly}\footnote{The reduced spectra and radial velocity observations presented in this work will be made available at the CDS ({\tt http://cdsweb.u-strasbg.fr/}) and at {\tt http://www.astro.keele.ac.uk/$\sim$jkt/}}. The single-Gaussian method gives radial velocities strongly affected by the peak of the emission line, so can highlight the velocity variation of the line emission of the bright spot on the edge of the accretion disc caused by the impact of material lost from the secondary star.

The double-Gaussian technique \citep{SchneiderYoung80apj} can be used to measure the velocity variation of the wings of emission lines. The flux in the line wings comes from material close to the inner edge of the accretion disc, so is generally interpreted as representing the motion of the white dwarf. While common, this interpretation is subject to systematic uncertainties \citep[see][]{Shafter83apj,Thorstensen00pasp}. The velocities are measured by cross-correlating the emission line spectrum against a function containing two Gaussians of equal FWHM and area but with one having positive height and one negative height. By choosing a small FWHM (similar to the resolution of the spectra), and a separation between the two Gaussians of slightly less than the FWHM of the emission line, the point at which the cross-correlation function is equal to zero represents the velocity of the line wings. In all cases we varied the FWHM and separation of the Gaussians to confirm the robustness of our results.

An independent orbital period determination can be carried out using the ratio of the fluxes in the redder and bluer parts of the line. Our procedure for this is to determine the wavelength where the line fluxes bluewards and redwards are equal, expressed as a velocity relative to the rest wavelength of the line. When reliable results were obtained from this method, they agreed in all cases with those we found from the radial velocities.

\begin{figure*} \includegraphics[width=\textwidth,angle=0]{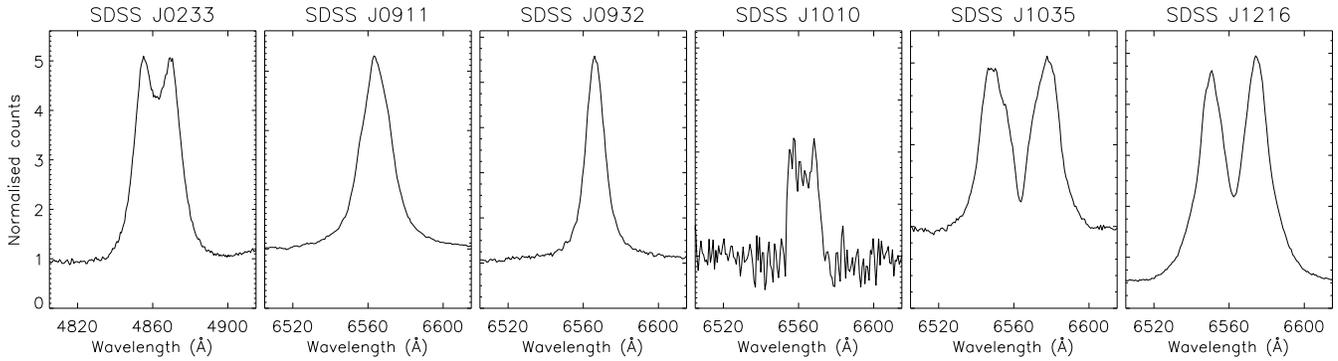} \\
\caption{\label{fig:avgspec} The averaged emission line profiles of the six CVs
for which we obtained spectra (H$\beta$ for SDSS\,J0233 and H$\alpha$ for the
others). The abscissa of each plot is in counts where the continuum level has been
normalised to unity.} \end{figure*}

\begin{figure} \includegraphics[width=0.48\textwidth,angle=0]{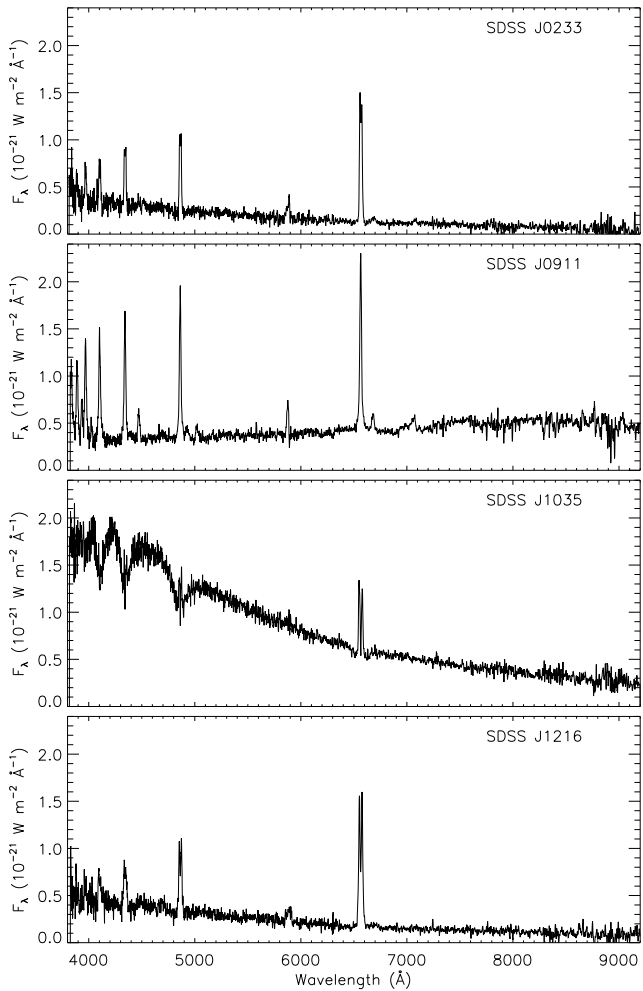} \\
\caption{\label{fig:sdssspec} The SDSS spectra of our four main targets, all
plotted using the same limits in wavelength and flux. The flux levels have been
smoothed with a 10-pixel Savitsky-Golay filter using the {\sc savgol} procedure
in {\sc IDL} (see {\tt http://www.rsinc.com/}). The units of the abscissa are
$10^{-21}$\,W\,m$^{-2}$\,\AA$^{-1}$, which corresponds to
$10^{-18}$\,erg\,s$^{-1}$\,cm$^{-2}$\,\AA$^{-1}$.}\end{figure}

Averaged emission line profiles are given in Fig.~\ref{fig:avgspec} and the SDSS spectra of the CVs for which we have measured orbital periods are shown in Fig.~\ref{fig:sdssspec} for reference.

\subsection{Orbital period measurement}                     \label{sec:data:period}

\begin{figure} \includegraphics[width=0.48\textwidth,angle=0]{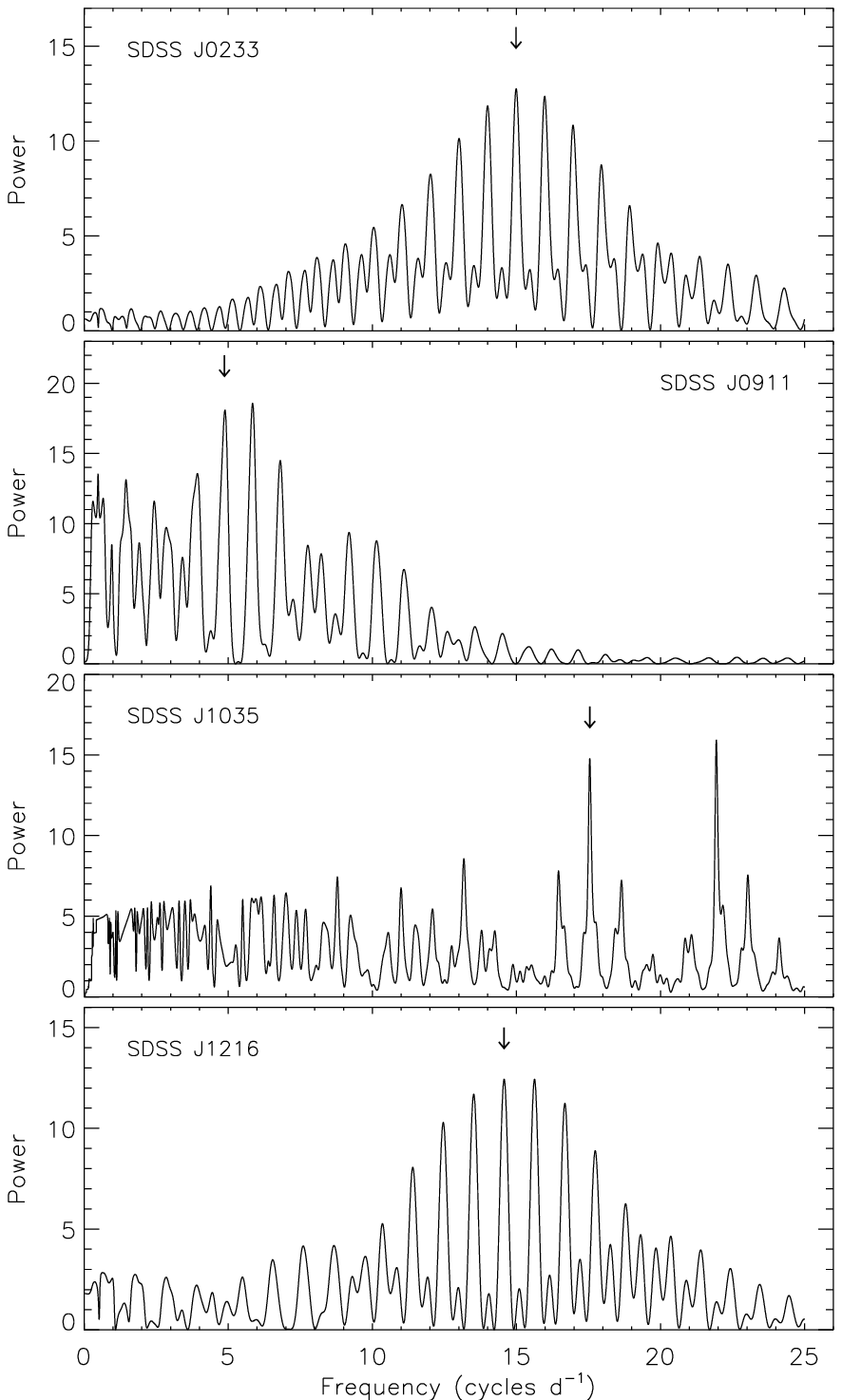} \\
\caption{\label{fig:period} Periodograms of the radial velocities measured for
SDSS\,J0233, SDSS\,J0911, SDSS\,J1035 and SDSS\,J1216. That for SDSS\,J1035 is
an ORT periodogram \citep{Schwarzenberg96apj} and the others are \citet{Scargle82apj}
periodograms. The frequencies corresponding to the best-fitting orbital periods from
{\sc sbop} are indicated with downward-pointing arrows.} \end{figure}

The measured radial velocities for each CV were searched for periods using periodograms computed using the \citet{Scargle82apj}, analysis of variance \citep[AoV; ][]{Schwarzenberg89mn} and orthogonal polynomial \citep[ORT; ][]{Schwarzenberg96apj} methods implemented within the {\sc tsa}\footnote{\scriptsize\tt http://www.eso.org/projects/esomidas/doc/user/98NOV/volb/node220.html} context in {\sc midas}. Periodograms for our four main targets are shown in Fig.~\ref{fig:period}.

We fitted single-lined circular spectroscopic orbits (equivalent to fitting sine curves) to the radial velocities using the {\sc sbop}\footnote{Spectroscopic Binary Orbit Program, written by P.\ B.\ Etzel, \\ {\tt http://mintaka.sdsu.edu/faculty/etzel/}} program, which we have previously found to give reliable error estimates for the optimised parameters \citep{Me+05mn}.

Due to the limited observing time available, we were not able to obtain data over a sufficient time or range of hour angles to avoid the periodograms containing strong power at the one-day aliases of the orbital periods of the CVs. In several cases this has made it difficult to be certain which of the possible peaks represents the true orbital period. When this problem has been encountered, we have fitted spectroscopic orbits to the radial velocities for each of the possible periods (Table~\ref{tab:aliases}). In general, we expect the orbit with the correct period to have both the lowest residuals of the fit, as the systematic deviations from the sine curve are minimised, and the highest velocity amplitude, as any mismatch in orbital phase would act to decrease the measured amplitude.

As a final check, we have also investigated the reliability of our period determinations using bootstrapping simulations, following the method outlined by \citet[][chapter 15.6]{Press+92book}. For each simulation we identified the point of highest power, which always fell close to one of the alias periods. The fraction of `successes' for each alias can then be interpreted as the probability that it represent the correct period. We would like to emphasise that this fraction generally represents an underestimate of the probability that the alias is the correct one. This is because it does not take into account the interactive rejection of aliases, normally because they result in clearly deformed radial velocity curves, which is possible when analysing the periodograms of the actual observational data for each system.

\begin{figure} \includegraphics[width=0.48\textwidth,angle=0]{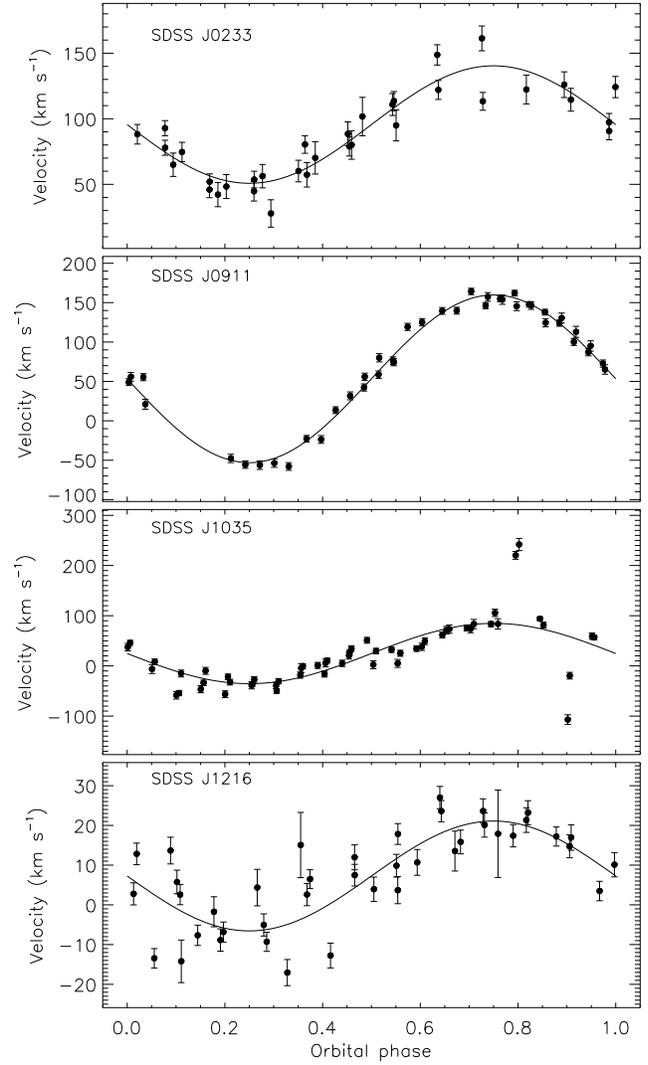} \\
\caption{\label{fig:orbit} Spectroscopic orbits for SDSS\,J0233, SDSS\,J0911,
SDSS\,J1035 and SDSS\,J1216.} \end{figure}

\begin{figure*}
\includegraphics[width=0.24\textwidth,angle=0]{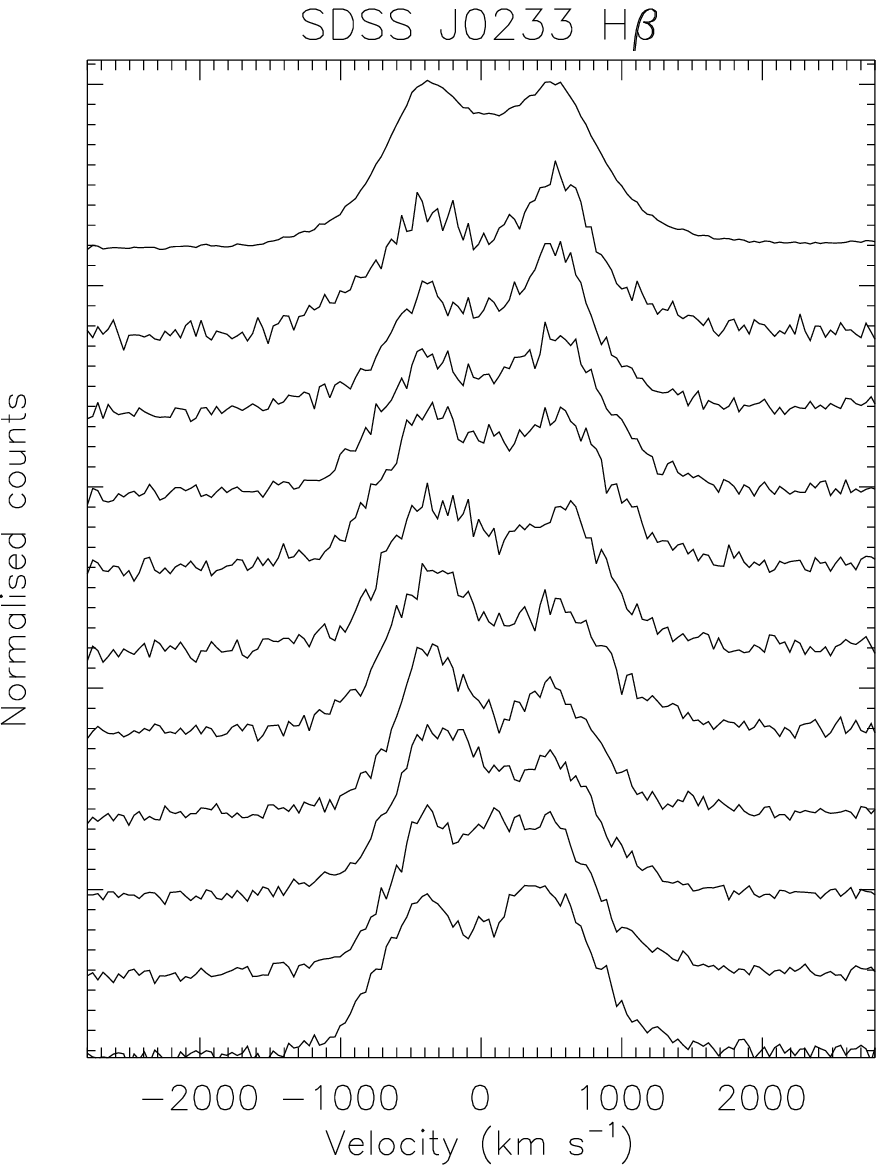}
\includegraphics[width=0.24\textwidth,angle=0]{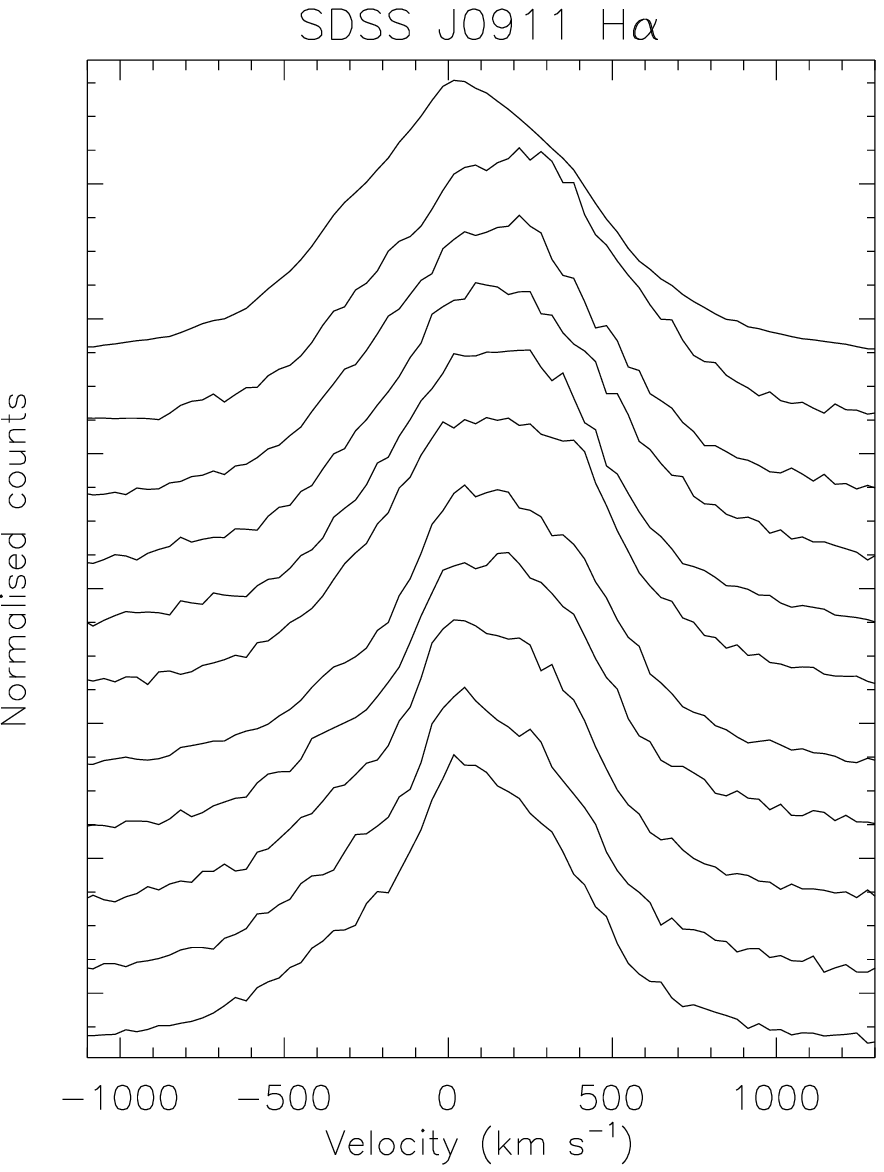}
\includegraphics[width=0.24\textwidth,angle=0]{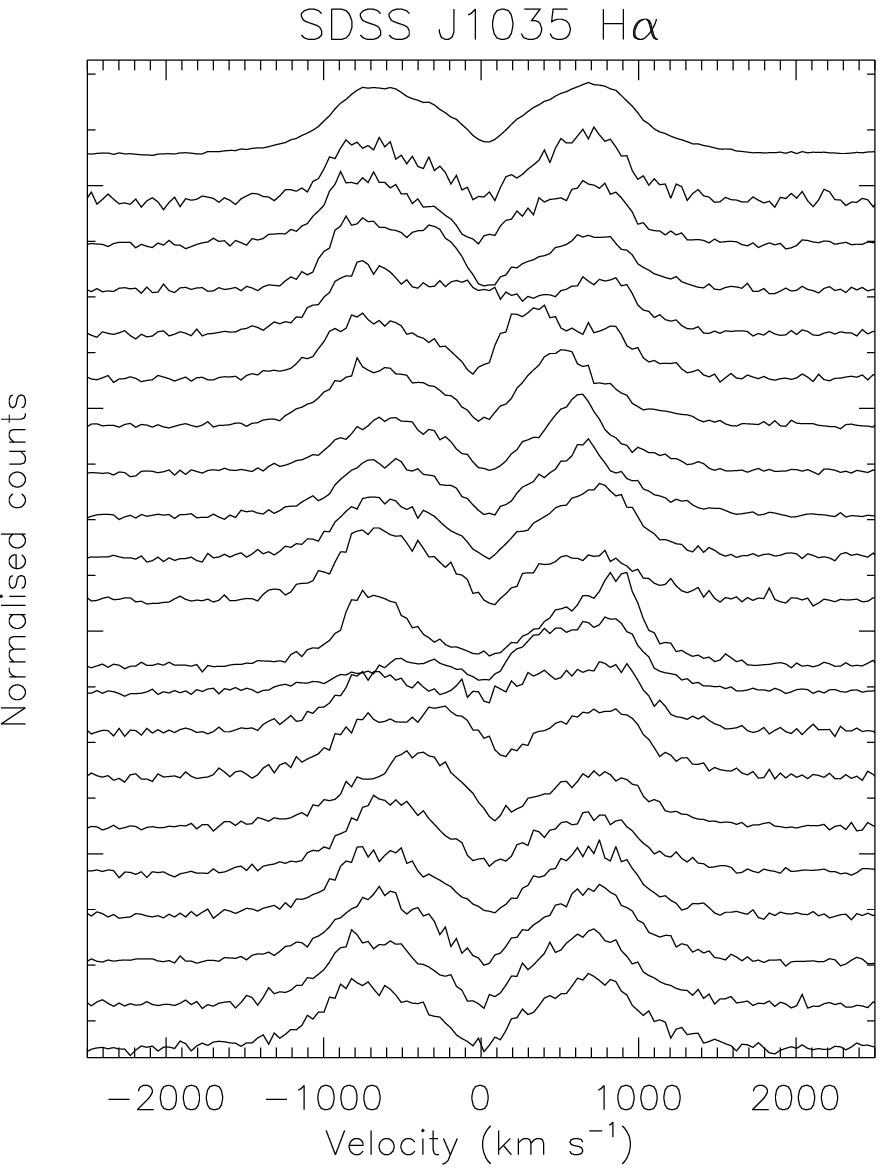}
\includegraphics[width=0.24\textwidth,angle=0]{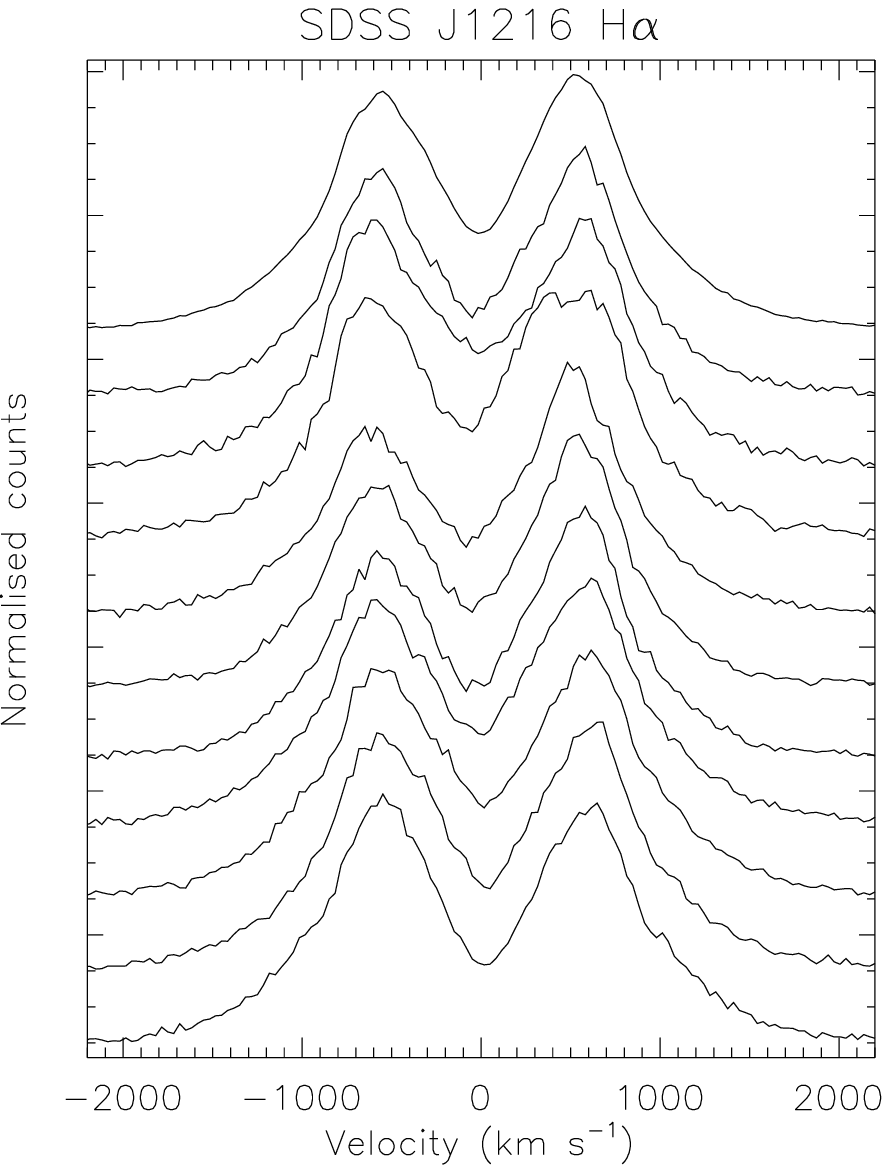}
\caption{\label{fig:stacked} Phase-binned stacked spectra of the
target CVs. The top spectrum in each plot is the averaged spectrum.}
\end{figure*}

\begin{figure*}
\includegraphics[width=0.24\textwidth,angle=0]{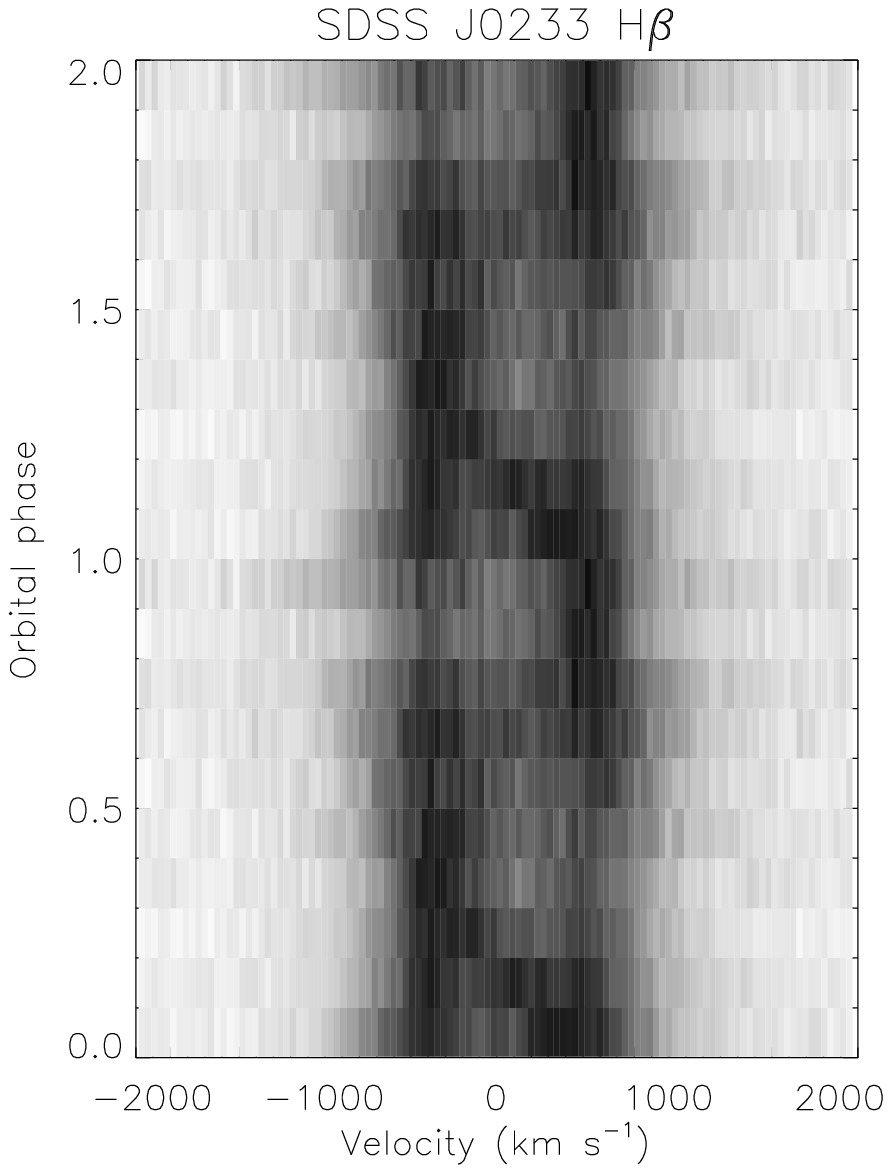}
\includegraphics[width=0.24\textwidth,angle=0]{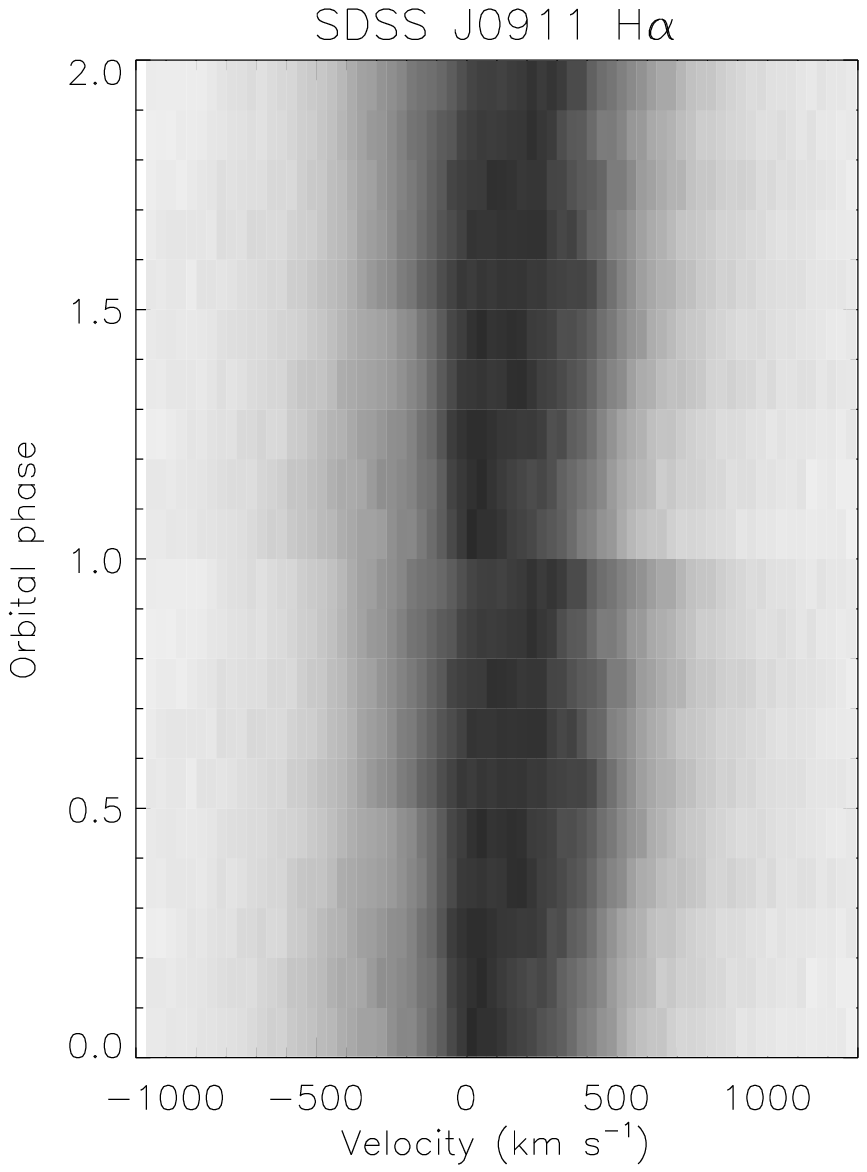}
\includegraphics[width=0.24\textwidth,angle=0]{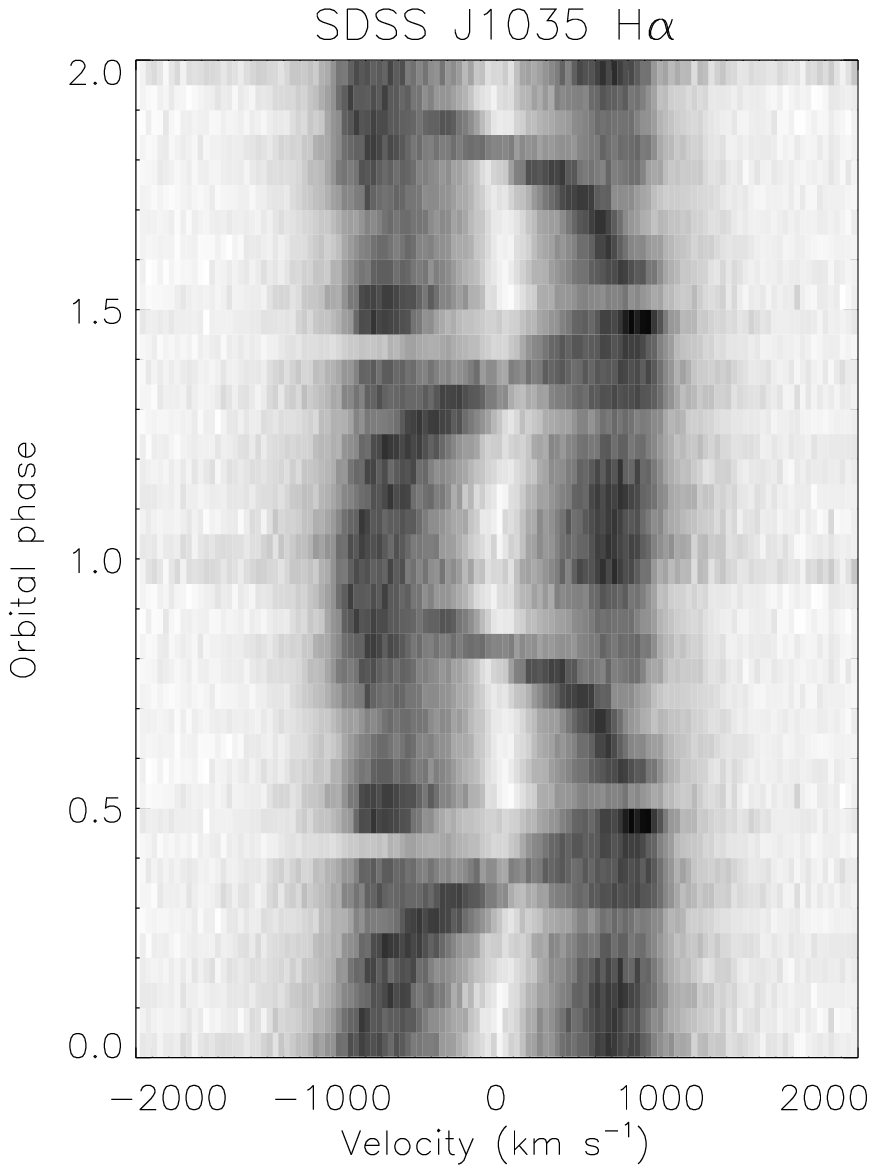}
\includegraphics[width=0.24\textwidth,angle=0]{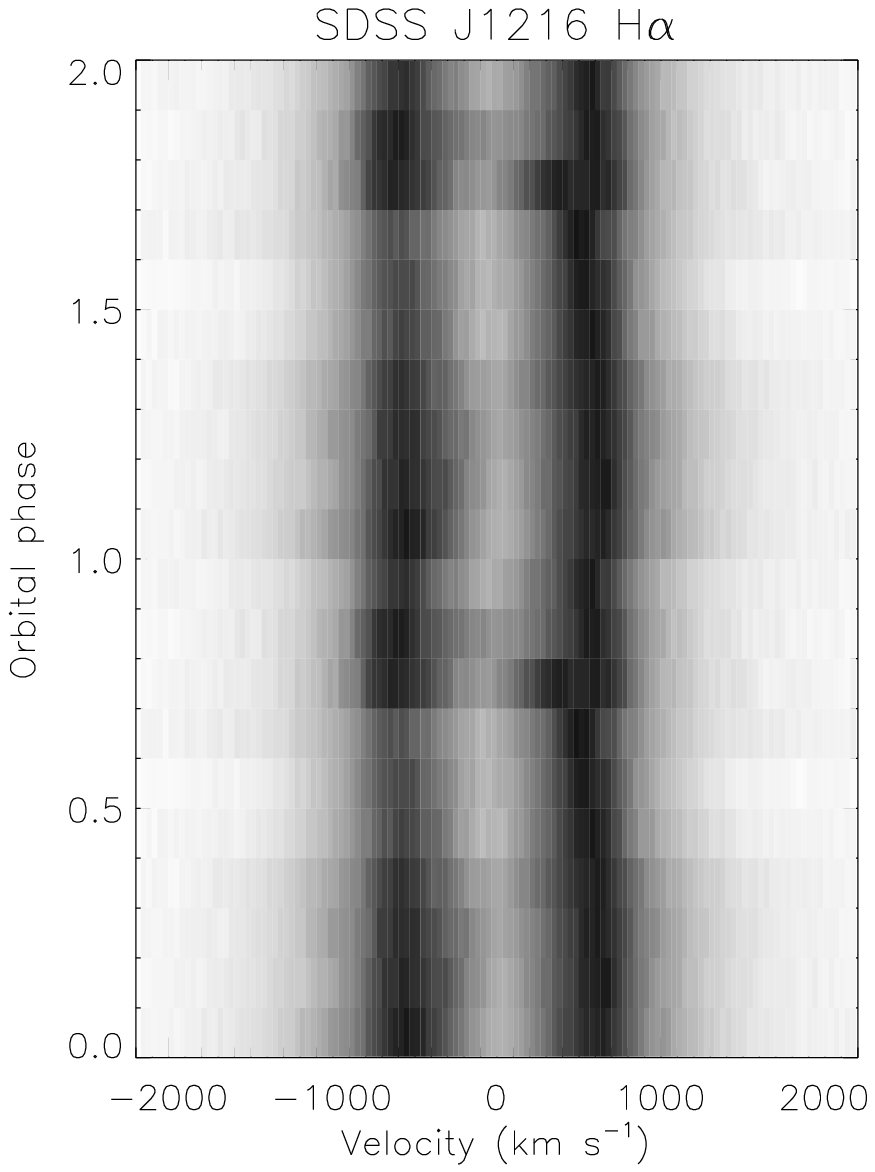}
\caption{\label{fig:trailed} Greyscale plots of the phase-binned
trailed spectra of the target CVs.} \end{figure*}

For our four main targets we show the spectroscopic orbits in Fig.~\ref{fig:orbit}, phased according to our preferred ephemeris for each system, and give the orbital parameters in Table~\ref{tab:orbits}. The corresponding phase-binned spectra are shown in Fig.~\ref{fig:stacked} (stacked) and Fig.~\ref{fig:trailed} (trailed).

\begin{table} \begin{center} \caption{\label{tab:aliases} Some spectroscopic
orbital parameters for the possible periods for SDSS\,J0233, SDSS\,J0911 and
SDSS\,J1216. $\sigma_{\rm rms}$ is the root mean square of the residuals of
the observations around the best fit. The adopted periods are shown using
bold type.}
\begin{tabular}{l c r@{\,$\pm$\,}l r} \hline
Target      &   Period        & \mc{Velocity amplitude}    & $\sigma_{\rm rms}$ \\
            &   (day)         & \mc{(\kms)}                &     (\kms)         \\ \hline
SDSS\,J0233 &      0.062692   &  42.2     & 4.0       & 15.2 \\
{\bf SDSS\,J0233} & {\bf 0.066724}  & {\bf 44.8}     &{\bf 3.4}       & {\bf12.7} \\
SDSS\,J0233 &      0.071446   &  44.3     & 3.7       & 13.7 \\ \hline
SDSS\,J0911 &      0.256545   & 124.4     & 4.0       & 12.8 \\
{\bf SDSS\,J0911} & {\bf 0.205374}  & {\bf 106.4}     &{\bf 2.3}       & {\bf 9.2} \\
SDSS\,J0911 &      0.171225   &  98.3     & 4.7       & 20.5 \\ \hline
SDSS\,J1216 &      0.059963   &  12.7     & 1.9       &  8.1 \\
SDSS\,J1216 &      0.064001   &  13.5     & 1.7       &  7.4 \\
{\bf SDSS\,J1216} & {\bf 0.068628}  & {\bf 13.8}     &{\bf 1.6}       & {\bf 7.0} \\
SDSS\,J1216 &      0.073986   &  13.7     & 1.7       &  7.3 \\
\hline \end{tabular}
\end{center} \end{table}

\begin{table*} \begin{center} \caption{\label{tab:orbits} Circular spectroscopic
orbits found using {\sc sbop}. The reference times (corresponding to zero phase)
refer to inferior conjuction (the midpoint of primary eclipse in eclipsing systems).}
\begin{tabular}{l r@{\,$\pm$\,}l r@{\,$\pm$\,}l r@{\,$\pm$\,}l r@{\,$\pm$\,}l r}\hline
Target&\mc{Orbital period}&\mc{Reference time}&\mc{Velocity amplitude}&\mc{Systemic velocity}&$\sigma_{\rm rms}$\\
      &     \mc{(day)}    &    \mc{(HJD)}     &       \mc{(\kms)}     &      \mc{(\kms)}     &      (\kms)      \\
\hline
SDSS\,J0233 &  0.066725 & 0.000064  &  53762.59652 & 0.00080  &   44.8 & 3.4  &  95.6 & 2.3  & {\it 12.7 } \\
SDSS\,J0911 &  0.205374 & 0.000152  &  53762.61735 & 0.00062  &  106.4 & 2.3  &  53.3 & 1.6  & {\it  9.2 } \\
SDSS\,J1035 &  0.056983 & 0.000061  &  53761.75152 & 0.00176  &   60.0 & 3.7  &  24.7 & 2.7  & {\it 18.2 } \\
SDSS\,J1216 &  0.068628 & 0.000112  &  53762.81177 & 0.00123  &   13.8 & 1.6  &   7.3 & 1.1  & {\it  7.0 } \\
\hline \end{tabular} \end{center} \end{table*}

\section{Results for each system}

\subsection{SDSS J023322.6+005059.5}

The discovery spectrum of SDSS\,J0233 was presented by \citet{Szkody+02aj} and shows slightly double-peaked Balmer emission lines along with weak He\,I emission (Fig.~\ref{fig:sdssspec}). \citet*{Woudt++04mn} obtained high-speed photometry of the system, finding both flaring and flickering activity but no clear indications of periodic behaviour.

The higher quality of the WHT spectroscopy compared to the SDSS data clearly shows that the Balmer lines are double-peaked, and suggests that SDSS\,J0233 is a disc-accreting system.  We were not able to extract a sensible radial velocity variation from the strong H$\alpha$ line, but a single-Gaussian approach to H$\beta$ indicated an orbital period of $\sim90$\,min. We therefore decided to obtain the VLT spectroscopy of SDSS\,J0233 centred on H$\beta$.

We have measured radial velocities from the H$\beta$ emission line in our VLT spectra using both single and double Gaussians. Radial velocities from the double Gaussian technique show a weak periodicity at 131.4 and 120.6 minutes, but those from single Gaussian measurements show a clear periodicity in the region of 96\,min (Fig.~\ref{fig:period}) which results in a sinusoidal velocity variation with orbital phase (Fig.~\ref{fig:orbit}). Trailed greyscale plots of the VLT spectra, phase-binned using this period, show that a weak S-wave is present. S-waves are commonly observed in CVs and are attributed to the velocity variation of a hot spot where the accretion stream from the secondary star hits the outer edge of the accretion disc. Because of the larger amplitude of the S-wave, compared to the wings of the emission lines, it is often a better indicator of orbital variation than the emission line wings.

The Scargle periodogram for SDSS\,J0233 (Fig.~\ref{fig:period}) and the ORT periodogram show the strongest power at 96\,min, but the two adjacent one-day aliases show only slightly lower power. In the AoV periodogram the 90\,min alias shows a much weaker peak but the 102\,min peak again has only slightly less power than the 96\,min highest peak. We have fitted spectroscopic orbits using all three possible periods and find that the 96\,min period gives the orbit with the lowest residuals and the highest velocity ampitude. Bootstrapping simulations using the Scargle periodogram method find that this alias has the highest power 75\% of the time (but remember that this is pessimistic -- see Sec.~\ref{sec:data:period}). We are therefore confident that this is the actual orbital period of the binary system. The corresponding best-fitting orbit has a period of $96.08 \pm 0.09$\,min.

\subsubsection{Light curve analysis for SDSS\,J0233}

\begin{figure} \includegraphics[width=0.48\textwidth]{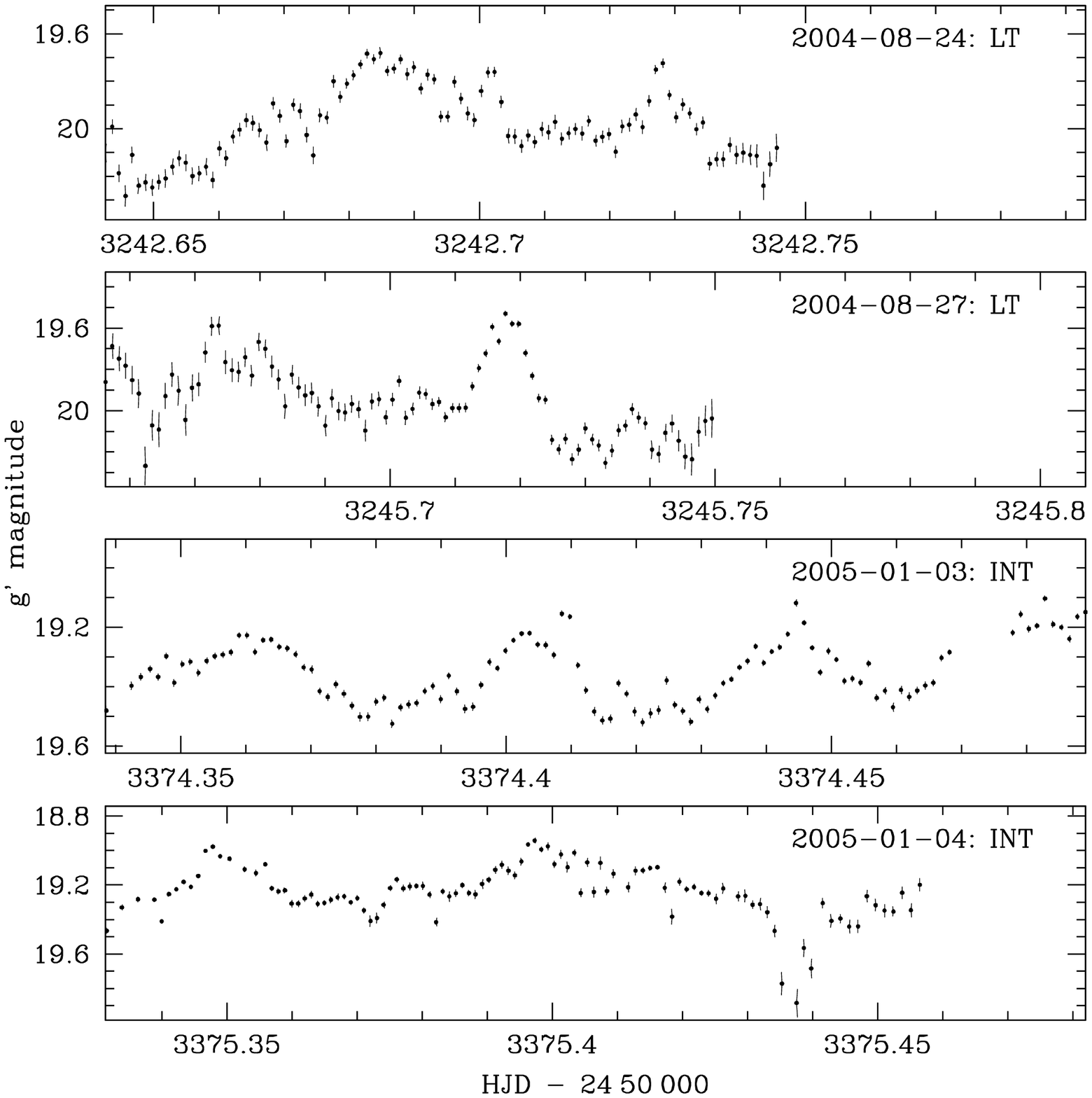} \\
\caption{\label{fig:0233lc} Sample light curves of SDSS\,J0233.} \end{figure}

\begin{figure*}
\includegraphics[angle=-90,width=0.48\textwidth]{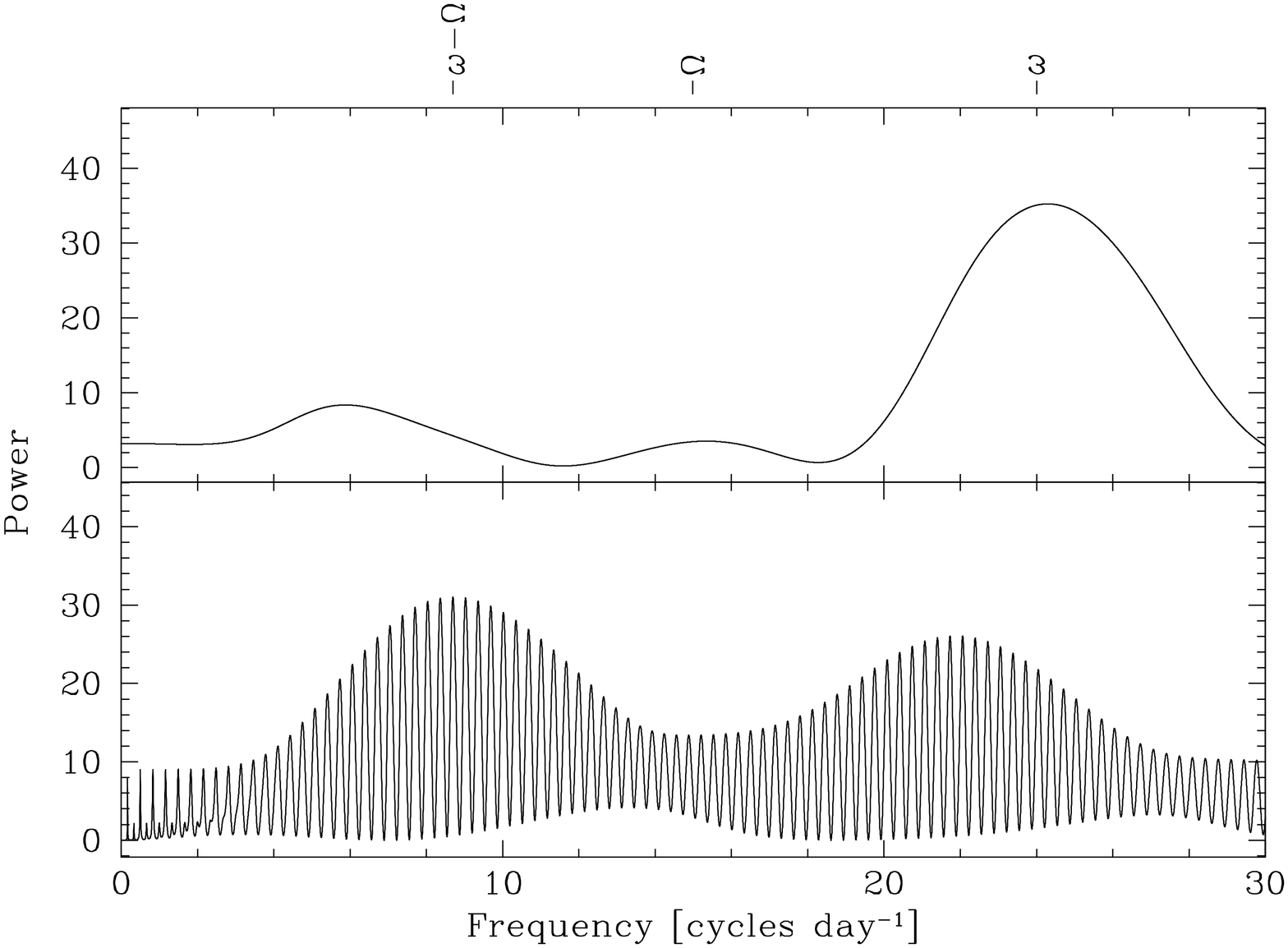}
\includegraphics[angle=-90,width=0.48\textwidth]{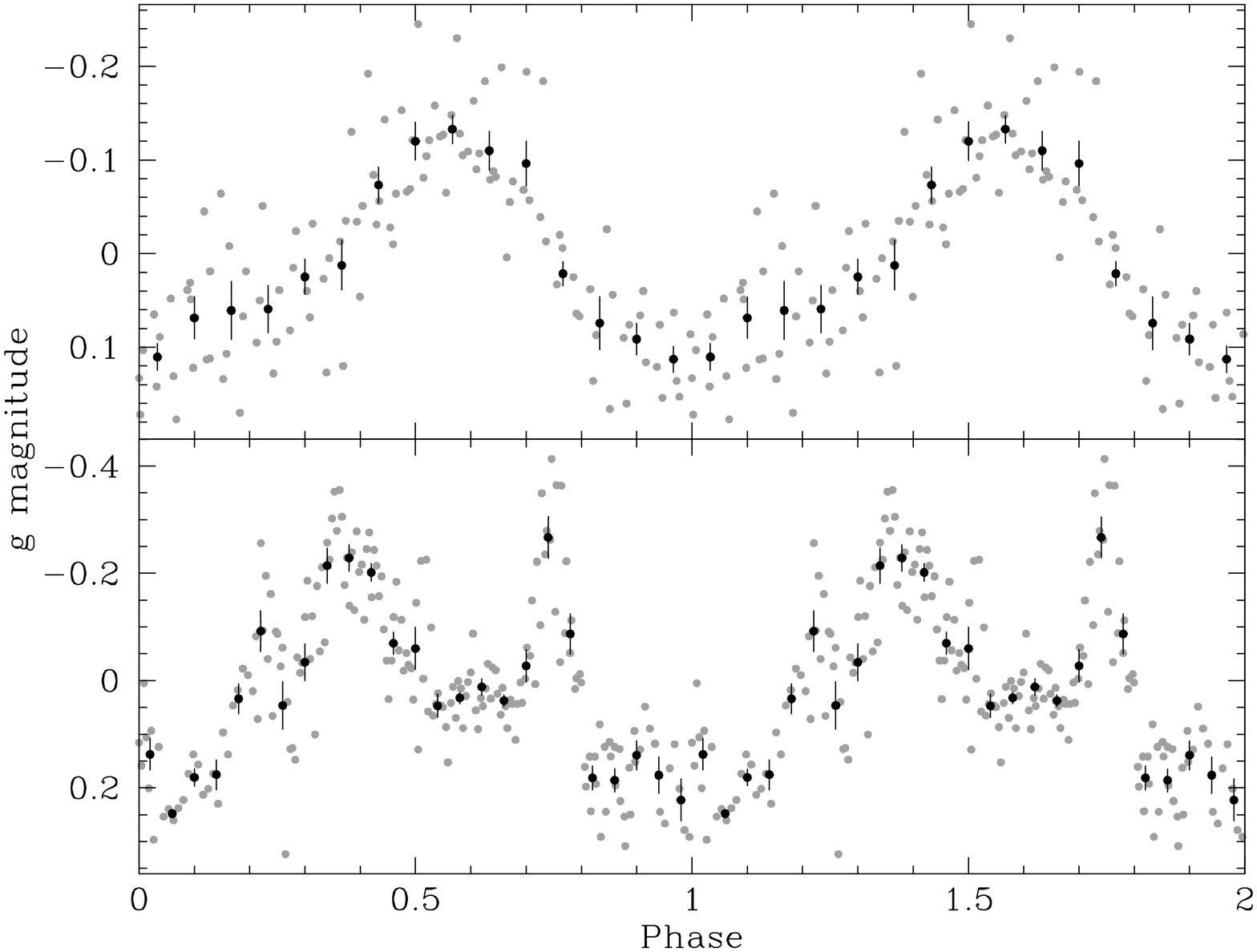}
\caption{\label{fig:0233tsa} Time series analysis of the LT and INT photometry
of SDSS\,J0233. Upper panels: The Scargle periodogram of the INT data of 2005
January 3. Left: a strong peak is present at 24\cd\ (60\,min). Right: the data folded over
that frequency (small points) and binned into 15 points. Lower panels: the
Scargle periodogram of the two combined nights of LT data from 2004 August. Left: the
strongest signal is found at 8.7\cd\ (166\,min), with a number of aliases spaced by
$\frac{1}{3}$\cd. Right: light curves of the original data folded with 8.7\cd\
(small points) and 25-point phase-binned data (large points).} \end{figure*}

The light curves of SDSS\,J0233 display a mixture of undulations on timescales of hours, superimposed by short-term flares lasting about 60\,min (Fig.~\ref{fig:0233lc}). These short-term brightenings were also detected in the photometry reported by \citet{Woudt++04mn}. Similar short events are seen on occasion as brightness depressions, e.g.\ on 2005 January 4 (Fig.~\ref{fig:0233lc}). A time series analysis of all data combined does not reveal any coherent period. However, different subsets of data show periodic variability. More specifically, a Scargle periodogram of the two nights of LT data obtained in 2004 August (Table~\ref{tab:obslog}) shows a strong signal near 160\,min (9\cd); the data folded over this period show a complex but clearly repeated structure (Fig.~\ref{fig:0233tsa}, lower panels). On 2005 January 3 a strong $\sim$60\,min modulation is seen in the data and picked up by a Scargle periodogram (Fig.~\ref{fig:0233tsa}, top). There is no indication of this period in the spectra. Given the spectroscopically determined period of 96\,min, we tentatively interpret the photometric signals as the white dwarf spin period of $\sim60$\,min and the beat between orbital and spin period of $\sim160$\,min. From this we suggest that SDSS\,J0233 is an intermediate polar with a low accretion rate. This is in agreement with the evidence for strong X-ray variability reported by \citet{Szkody+02aj}.

\subsection{SDSS J091127.36+084140.7}

\citet{Szkody+05aj} presented the discovery of this CV on the basis of an SDSS spectrum which shows strong hydrogen Balmer emission and some \ion{He}{i} line emission superimposed on a slightly red continuum (Fig.~\ref{fig:sdssspec}).

We obtained a total of 42 H$\alpha$ spectra from three nights of VLT observations. Radial velocities were measured using the double Gaussian technique with Gaussian FWHMs of 300\kms\ and separation of 1500\kms. All three types of periodogram show strong power at periods close to 295\,min. The one-day alias at 245\,min has the strongest power in the Scargle periodogram, but phasing the radial velocities with this period gives a discontinuous radial velocity curve which suggests that the Scargle method does not perform as well when the phase coverage is incomplete. The AoV and ORT alternatives show only weak power at this period but power at the other one-day alias, 369\,min, which is nearly as strong as for the 295\,min period. Inspection of the phased data shows that the 295\,min period gives nice sinusoidal radial velocity variations whereas the 369\,min alias again result in a clearly deformed radial velocity curve. We are confident that an orbital period of close to five hours is correct, and fitting the radial velocities with {\sc sbop} gives a spectroscopic orbit with a period of $295.74 \pm 0.22$\,min.

At the suggestion of the referee, we have compared the red part of the SDSS spectrum of SDSS\,J0911 to spectral templates of M-dwarf stars in order to determine the spectral type of the secondary component. We find M2$\pm$0.5 from the relative strengths of the visible molecular absorption bands. Using the calibration of the surface brightness in the 7165\,\AA\ and 7500\,\AA\ TiO bands by \citet{BeuermannWeichhold99conf}, and assuming a Roche-lobe-filling mass donor with radius $R_2 = (3.7\!\pm\!0.2) \times 10^{8}$\,m, we estimate the distance to SDSS\,J0911 to be $d = 2000 \pm 300$\,pc. This distance and the galactic latitude of the system, $b = +35^{\circ}$, imply that it is located $\sim1.1$\,kpc above the galactic plane.


\subsection{SDSS J093238.2+010902.5}

\begin{figure} \includegraphics[width=0.4\textwidth,angle=0]{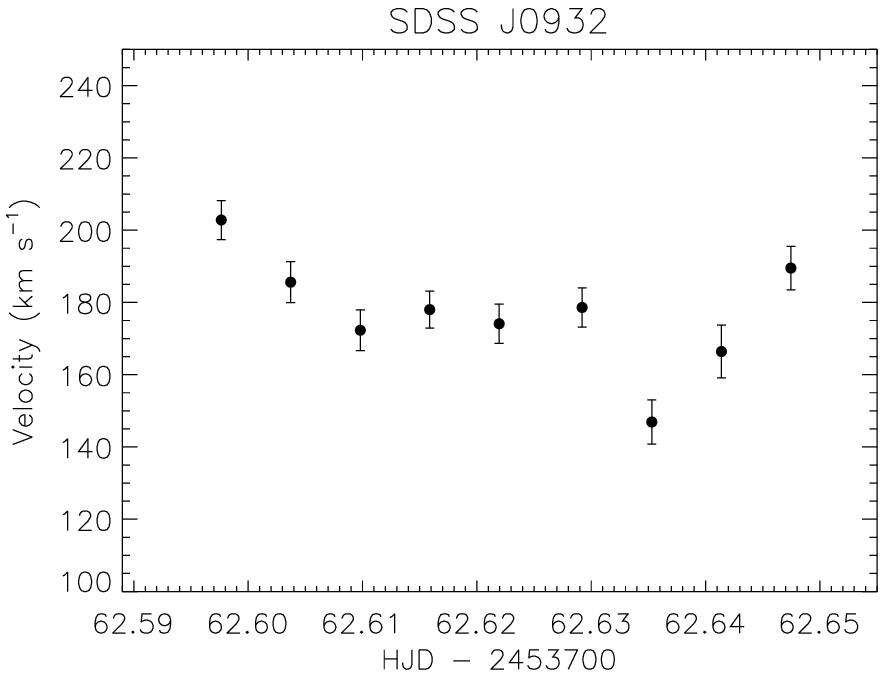} \\
\caption{\label{fig:0932rv} Measured H$\alpha$ radial velocities of SDSS\,J0932.}
\end{figure}

SDSS\,J0932 was discovered to be a CV by \citet{Szkody+03aj} from its SDSS spectrum, which shows a faint flat continuum with reasonably strong hydrogen emission lines and slight \ion{He}{i} emission.

We obtained a total of nine VLT spectra of SDSS\,J0932 over an 80\,min time period. These show H$\alpha$ emission with a FWHM of about 900\kms\ (Fig.~\ref{fig:avgspec}), which is smaller than in the four systems for which we have measured radial velocity variations. The \ion{He}{i} 6678\,\AA\ line is also clearly in emission, with an equivalent width of about one sixth of H$\alpha$. Our preliminary reductions at the telescope gave no clear indication of radial velocity changes due to orbital motion, so no further observations were obtained. Fig.~\ref{fig:0932rv} shows radial velocities measured using the double Gaussian technique with a FWHM of 300\kms\ and separation of 1000\kms; similar results are obtained for a range of values for Gaussian FWHM and separation.

It is certainly possible that this is a system with an orbital period much longer than the length of our dataset, which was observed over an interval in its orbit when its radial velocity was changing by very little. If this is the case then our survey will be slightly biassed against longer-period CVs, so we intend to revisit SDSS\,J0932 at a later date to further investigate this system.


\subsection{SDSS J101037.05+024914.9}

SDSS\,J1010 was identified as a CV by \citet{Szkody+03aj} from its SDSS spectrum, which shows a faint flat continuum with exceptionally strong hydrogen emission lines and faint \ion{He}{i} line emission.

The apparent magnitude of SDSS\,J1010 in both its SDSS imaging and spectroscopic observations was around 20.5, which is well within reach of VLT/FORS2 spectroscopy using our observational setup. However, the five spectra we obtained contained about ten times fewer counts than expected, indicating that the object was much fainter than during the SDSS observations. The average spectrum is shown in Fig~\ref{fig:avgspec} and displays weak H$\alpha$ emission.

Once photometry on our acquisition image confirmed that SDSS\,J1010 was dimmer than expected, we obtained images in the $B$ and $V$ passbands. The magnitudes of SDSS\,J1010 in these images is close to 22.0, confirming that SDSS\,J1010 was in a lower state at the time of our observations than those of the SDSS photometry and spectroscopy.

We have measured the equivalent width of the H$\alpha$ line to be approximately 600\,\AA\ in the SDSS spectrum. This is exceptionally large even for a short-period dwarf nova and suggests that the continuum is very weak, perhaps due to a cold white dwarf and a very late-type secondary star. The equivalent width of H$\alpha$ in the average spectrum from our VLT data is only $43 \pm 5$\,\AA, so the emission has weakened substantially in the interval between the SDSS and our VLT observations. Further observations of SDSS\,J1010 are desirable, as this object may be an example of the low-luminosity evolved CVs whose existence is predicted by theoretical models of CV evolution but which have not yet been observationally confirmed.


\subsection{SDSS J103533.02+055158.3}

SDSS\,J1035 is a particularly interesting system: its SDSS spectrum \citep{Szkody+06aj} shows relatively weak but double-peaked Balmer emission lines superimposed on a very blue continuum with unmistakable wide Balmer absorption lines coming from the white dwarf in the system (Fig.~\ref{fig:sdssspec}).

The brightness of SDSS\,J1035 allowed us to reduce our VLT/FORS2 exposure times to 200\,s, resulting in a total of 58 spectra. These were measured for radial velocity, with best results obtained using double Gaussians separated by 2000\kms\ and with FWHMs of 300\kms. The disappearance of the blue or red emission line components in certain observations (Figs.\ \ref{fig:stacked} and \ref{fig:trailed}) indicates that SDSS\,J1035 is an eclipsing system. The changing line shapes through eclipse have strongly affected a few of the radial velocities, so we searched for periods using the AoV method. We find a period of $82.06 \pm 0.09$\,min (Fig.~\ref{fig:period}) from fitting the radial velocities with {\sc sbop}. There is an additional possible period at 65\,min, where the two observed eclipses occur at the same phase, but this gives a clearly poorer radial velocity curve and so is very unlikely to reflect the actual period of the system. The 82\,min period is confirmed by subsequent photometry (Littlefair et al., submitted). For this period, the eclipses occur at phase 0.85 in the emission-line spectroscopic orbit, which is quite distant from the expected position at phase zero for a Keplerian orbit. This effect has been noted in quite a few CVs and indicates that radial velocities obtained from emission line wings are not always a good indicator of the motion of the white dwarf \citep{Thorstensen00pasp}.

\subsubsection{A model for the optical spectrum of SDSS\,J1035}

Following the approach described in \citet{Gansicke+99aa}, \citet{Rodriguez+05aa} and \citet{Gansicke+06mn}, we have fitted a three-component model to the SDSS spectrum of SDSS\,J1035. The three components are a synthetic pure hydrogen white dwarf spectrum calculated with the {\sc tlusty} and {\sc synspec} codes \citep{HubenyLanz95apj}, an isothermal and isobaric hydrogen disc \citep*{Gansicke++97mn} and an observed late-type spectral template (M0.5--M9 from \citealt{Beuermann+98aa} and L0--L8 from \citealt{Kirkpatrick+99apj}). We fixed the surface gravity of the white dwarf model spectra to $\logg = 8.0$, corresponding to $M_{\rm WD} \approx 0.6$\Msun, and a radius of $R_{\rm WD} \approx 8.7 \times 10^8$\,cm (assuming a carbon-core mass-radius relation from \citealt{HamadaSalpeter61apj}). The radius of the secondary star is constrained by the size of the secondary Roche lobe, which primarily depends on the orbital period, and we fixed $R_2 = 8.5 \times 10^{9}$\,cm. Free parameters of the three-component model are the white dwarf temperature, $T_{\rm WD}$, the spectral type of the secondary star, SpT(2), the temperature $T_{\rm disc}$ and the column density $\Sigma_{\rm disc}$ of the hydrogen slab, and the distance, $d$, to the system used to scale all three components.

\begin{figure} \includegraphics[angle=-90,width=\columnwidth]{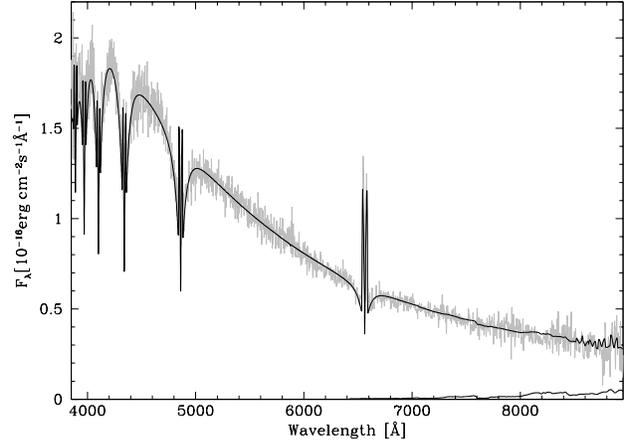} \\
\caption{\label{fig:1035sed} Three-component model of the SDSS spectrum
of SDSS\,1035. The three components are a white dwarf with $T_{\rm WD} =
11\,700$\,K, $R_{\rm WD} = 8.7 \times 10^8$\,cm; and isothermal and
isobaric hydrogen slab with $T_{\rm disc} = 6500$\,K and $\Sigma_{\rm disc}
= 5.3 \times 10^{-2}$\,g\,cm$^{-2}$, and an L0 secondary star, all scaled
to a distance of $d = 280$\,pc.} \end{figure}

A plausible model of the SDSS spectrum of SDSS\,J1035 is achieved for the following set of parameters (Fig.~\ref{fig:1035sed}): $T_{\rm WD} = 11\,700$\,K, SpT(2)\,=\,L0, $T_{\rm disc} = 6500$\,K, $\Sigma_{\rm disc} = 5.3\times10^{-2}$\,g\,cm$^{-2}$, and $d = 280$\,pc. The radius of the accretion disc implied by the flux scaling factor, at a distance of 280\,pc and assuming an inclination of $85^\circ$, is $R_{\rm disc} \approx 2 \times 10^{10}$\,cm, well within the Roche lobe of the primary. The spectral type of the secondary star is constrained by the non-detection of TiO bands in the red part of the spectrum of SDSS\,J1035 and represents only an ``early'' limit -- the true spectral type could well be later than L0. The white dwarf contribution to the optical spectrum of SDSS\,J1035 is approximately 85\%.

The parameters of the white dwarf are subject to uncertainty in its mass, which cannot be constrained from the optical spectrum alone, and to the contamination of the Balmer absorption lines by emission from the accretion disc. We estimate the uncertainties in the temperature of the white dwarf and distance to be approximately $\pm$1500\,K and $\pm$60\,pc, respectively. Adopting a larger surface gravity of $\logg = 9.0$ for the white dwarf (corresponding to a mass of 1.2\Msun) causes the best-fit parameters for the spectrum of SDSS\,J1035 to change to $T_{\rm WD} = 14\,000$\,K, $d = 140$\,pc, and an upper ``early'' limit on  the secondary spectral type of L2.
 
Comparing the properties of SDSS\,J1035 to those of SDSS\,J133941.11+484727.5 \citep{Gansicke+06mn}, it appears that the systems are extremely similar, with orbital periods close to the observed 80\,min period minimum and relatively cool white dwarfs which nevertheless dominate the optical spectrum of the systems. The implications are that both systems have very low mass transfer rates, and hence low-luminosity accretion discs and bright spots. They are good candiates for brown-dwarf-mass donor stars, as even the presence of a late-type M dwarf is difficult to reconcile with the non-detection of TiO absorption bands in the red part of the spectrum. Both systems thus have in all aspects the properties predicted by CV evolution theory for old CVs that have evolved to, or past, the minimum period.

\subsubsection{Doppler tomography of the accretion disc of SDSS\,J1035}

The prominent S-wave visible in SDSS\,J1035 (Fig.~\ref{fig:trailed}) is comparable to that of WZ\,Sge \citep{SpruitRutten98mn}. There is dynamical information in such S-waves which we attempted to extract by application of the method of Doppler tomography \citep{MarshHorne88mn}. In this procedure the spectra are modelled as the sum of a series of sinusoids which have radial velocities as a function of orbital phase $\phi$ obeying the relation
\[ V(\phi) = - V_X \cos 2\pi \phi + V_Y \sin 2\pi \phi \]
A Doppler map is a plot of the intensity of each sinusoid as a function of $V_X$ and $V_Y$. We applied the maximum entropy-based method of \citet{MarshHorne88mn}. A complication in the case of SDSS\,J1035 is the deep central absorption clearly visible in Fig.~\ref{fig:trailed}. This is lower than any emission pattern over the disc can produce, and is presumably the result of the underlying white dwarf photospheric absorption line with possibly some contribution from absorption through the disc along the line of sight to the white dwarf. This causes a slight difficulty when applying the maximum entropy method, which assumes positive intensities at all points. Here we sidestep this problem by first adding Gaussians of FWHM 600\kms\ to all the profiles of sufficient strength to allow the corresponding Doppler map to be positive. We then computed the map (a two-dimensional Gaussian) corresponding to the Gaussians and subtracted this from the maps. This procedure does not affect the form of the map because we use an entropy that is only sensitive to structure on small scales (less than 250\kms).

\begin{figure} \includegraphics[angle=270,width=\columnwidth]{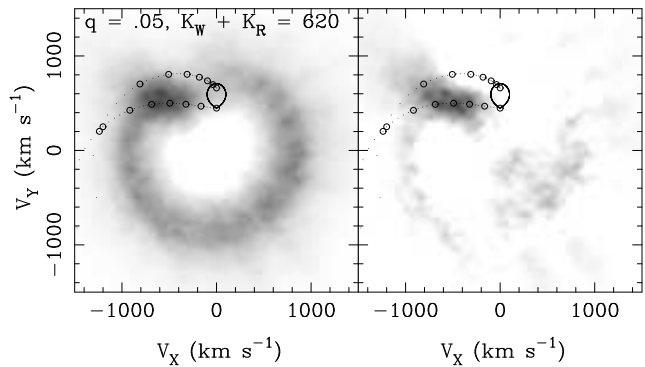} \\
\caption{\label{fig:1035map} The Doppler image in H$\alpha$ of SDSS\,J1035.
The Roche lobe of the mass donor and tracks of the gas stream for the
parameters indicated are shown. The lower track is the direct velocity of the
stream while the upper represents the Keplerian velocity of the disc along the
path of the stream. The left-hand panel shows the Doppler image whilst the
right-hand panel shows the Doppler image after the subtraction of an azimuthal
average to emphasise the asymmetries.} \end{figure}

\begin{figure} \includegraphics[angle=270,width=\columnwidth]{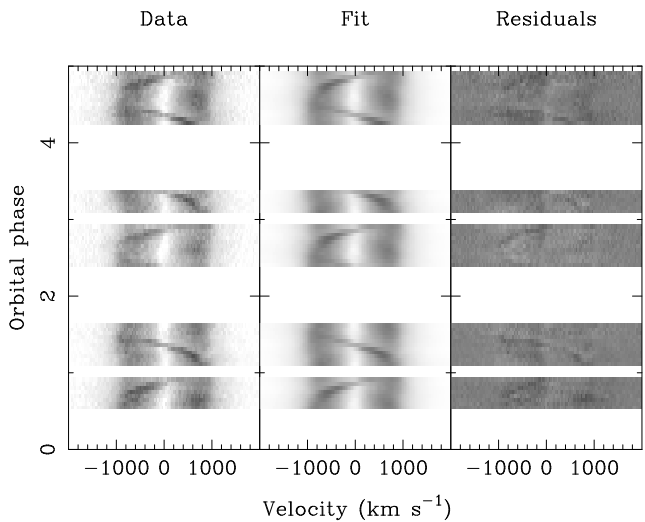} \\
\caption{\label{fig:1035trail} The H$\alpha$ profiles of SDSS\,J1035 are shown
on the left, with integer cycle gaps removed. The middle panel shows the fit
based upon the map of Fig.~\protect\ref{fig:1035map}, while the right-hand
panel shows the difference between the two. The greyscales of left-hand and
middle panels run from 0 to 5 while the right-hand panel is scaled from $-2.5$
to $+2.5$, in units of the continuum level.} \end{figure}

The maps were computed avoiding phases from $-0.08$ to $+0.08$ to avoid spectra affected by the eclipse, as eclipses are not accounted for in standard Doppler tomography. We used the ephemeris
\[ T_0({\rm HJD}) = 2453799.4807237 + 0.0570076 \times E \]
derived from independent photometric observations (Littlefair et al., submitted). The map is shown in the left-hand panel of Fig.~\ref{fig:1035map}; the right-hand panel shows the result of subtraction of an azimuthally averaged version of the map to bring out the asymmetries. Fig.~\ref{fig:1035trail} shows the spectra, fit and residuals. We plot a representative path of the gas stream that goes through the bright spot. Equally good fits can be obtained for many values of the mass ratio, $q$, and the sum of the velocity amplitudes of the two stars, $K_{\rm WD}$+$K_{\rm 2}$ (\kms). Examples are ($q$,$K_{\rm WD}$+$K_{\rm 2}$) = $(0.02,550)$, $(0.03,590)$, $(0.05,620)$ and $(0.07,650)$, although the last pair imply a white dwarf mass in excess of the Chandrasekhar limit. It is not, however, certain that the bright-spot emission tracks the velocity of the gas stream; indeed the map here is reminiscent of the Doppler map of U\,Gem for which the emission is thought to lie between the two tracks plotted here \citep{Marsh+90apj,Unda++06mn}. This would reduce the value of $K_{\rm WD}$+$K_{\rm 2}$ necessary to match the bright spot and hence relax the constraint upon the mass ratio implied by the Chandrasekhar limit. One final noteworthy point is the velocity of the ring from the outer disc visible in the map which peaks at a radius of about 800\kms. This is one of the highest in our experience; it suggests a combination of a fairly massive white dwarf and/or a small disc.


\subsection{SDSS J121607.03+052013.9}

\begin{figure*} \includegraphics[width=0.9\textwidth,angle=0]{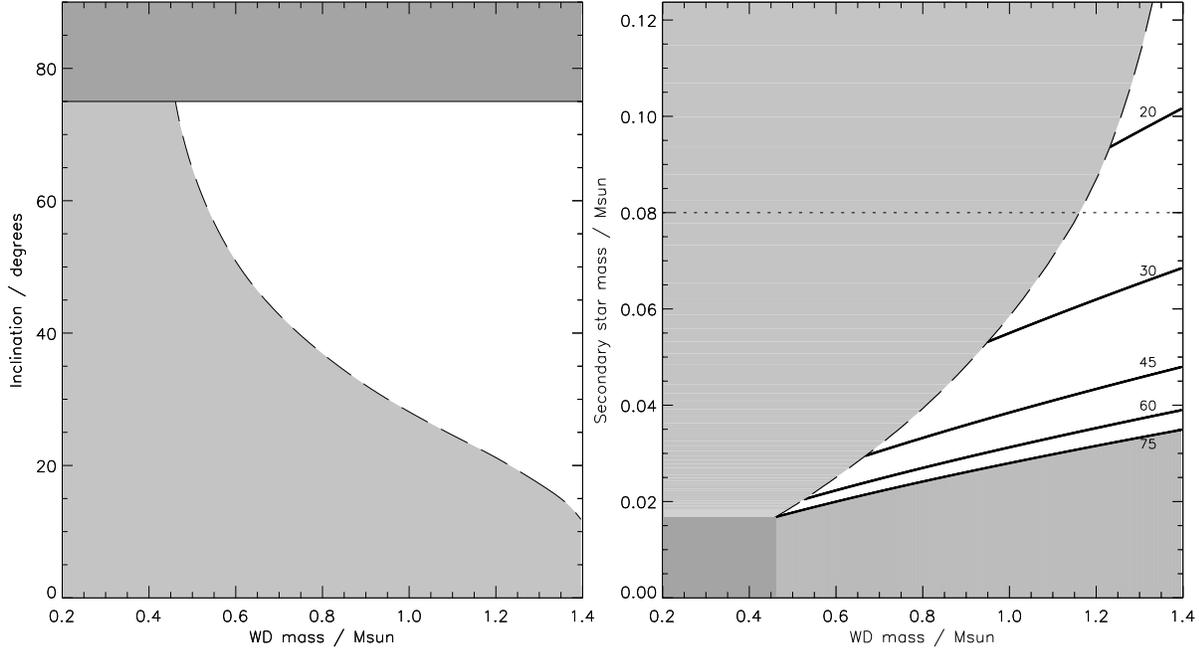} \\
\caption{\label{fig:1216mass} Diagrams of white dwarf mass versus orbital
inclination (left) and secondary star mass (right) for SDSS\,J1216 (see
Section~\ref{sec:1216:constraints} for further explanation). The dark shaded
regions indicate areas where inclination must be 75$^\circ$ or more (which is not
expected). The light shaded regions show where an accretion disc on a Keplerian
orbit at the observed velocity of the edges of the H$\alpha$ emission line would
be inside the surface of the white dwarf, which is not possible. The dashed lines
indicate the edge of this region. The solid lines indicate the locus of points
satisfying the mass function for orbital inclinations of 75$^\circ$, 60$^\circ$,
45$^\circ$ and 30$^\circ$ (labelled). The canonical minimum mass for core hydrogen 
burning is shown using a dotted line.} \end{figure*}

SDSS\,J1216 was discovered to be a CV by \citet{Szkody+04aj} from its SDSS spectrum, which shows a faint flat continuum with double-peaked hydrogen emission lines exhibiting a strong Balmer decrement. There is also faint He\,I emission which is strongest in the 5876\,\AA\ line (Fig.~\ref{fig:sdssspec}).

We measured radial velocities of the H$\alpha$ line in our 41 VLT spectra using both double Gaussians (separation 2000\kms\ and FWHM 300\kms\ provided the best results) and a single Gaussian function (FWHM 2000\kms). The Scargle periodogram of the resulting radial velocities contains a strong signal at two aliases, with almost equal power, at 92\,min and 99\,min. The two adjacent aliases of 86\,min and 107\,min also have almost as strong power. The AoV periodogram clearly prefers the 99\,min period and has only weak power at 106.5\,min. The ORT periodogram, however, prefers 92\,min, with the second-highest power being at 99\,min. On the basis of the periodograms alone, therefore, we have identified the 99\,min period as most the likely to represent the orbital motion, but cannot fully rule out the 92\,min period. Bootstrapping simulations confirm that 99\,min is the most likely period but that there is a significant probability that 92\,min (30\% for the Scargle and ORT periodograms, 10\% for AoV) or 107\,min (about 20\% for the three periodigram types) is the correct one.

We have fitted spectroscopic orbits to the radial velocities using the four possible periods, finding that the 99\,min alternative gives both the smallest residuals and the largest velocity amplitude, with the 107\,min option marginally second-best in both criteria. As our methods generally prefer the 99\,min period, with either the longer or shorter one-day aliases of this being the second-best option, we expect that this period is the correct one. We have fitted the radial velocities with {\sc sbop} to find a period of $98.82 \pm 0.16$\,min, resulting in a radial velocity curve which is sinusoidal. This period is also reproduced in studies of the ratio of the fluxes in the blue and red parts of the emission lines (see Sec.~\ref{sec:data:rv}). However, given that we have not been able to fully dismiss the alternative periods either side of 99\,min, and that this object has other particularly interesting properties, a futher study of SDSS\,J1216 would be desirable.

\subsubsection{Nature of the donor star in SDSS\,J1216}       \label{sec:1216:constraints}

The spectroscopic orbit we have measured for SDSS\,J1216 (Table~\ref{tab:orbits}) has a very small velocity amplitude of $13.8 \pm 1.6$\kms, despite the presence of double-peaked line emission suggesting a high inclination, indicating that the system has an extreme mass ratio. We can apply several constraints on the possible properties of the component stars in the system (see Fig.~\ref{fig:1216mass}):--%

\begin{itemize}
\item We have assumed in this discussion that the velocity amplitude found from the line wings is representative of the actual orbital motion of the white dwarf. This assumption is likely to be only approximately true, but the emission line motion is generally accepted to overestimate the actual movement of the white dwarf, so this is a conservative assumption for the discussion below. It would be useful to measure the WD absorption features directly, which may be possible in the future from ultraviolet spectroscopy of the system.%

\item A relation between masses and radii of white dwarfs was obtained (P.\ Bergeron, private communication) assuming $\Teff = 15\,000$\,K, supplemented with the zero-temperature predictions of \citet{HamadaSalpeter61apj} from 1.2\Msun\ to 1.4\Msun. Our use of two different sources and our choice of \Teff\ makes a negligible difference to the results found below.%

\item The full width at zero intensity of emission of the H$\alpha$ line gives a lower limit on the mass of the white dwarf. This quantity is notoriously tricky to measure for wide emission lines of faint stars (\citealt{Warner95book}) so we constructed a summed H$\alpha$ spectrum in which the (very small) orbital motion of the line wings was removed. The full width was measured to be $100 \pm 10$\,\AA, corresponding to a half-width zero-intensity velocity of $2290 \pm 230$\kms. Assuming the mass--radius relation for white dwarfs we adopted above, this velocity gives the orbital inclination as a function of mass for which the innermost part of the disc is on a Keplerian orbit at the surface of the white dwarf. This provides an upper limit on the white dwarf radius which translates into a lower limit on its mass.%

\item The spectra show no sign of eclipses (which can be seen by comparing the trailed spectra of SDSS\,J1216 with those of SDSS\,J1035 in Fig.~\ref{fig:trailed}), which indicates that the orbital inclination is below about 75$^\circ$ (\citealt{Hellier01book}).%

\item The mass function of the orbit is $(1.8 \pm 0.6) \times 10^{-5}$\Msun. For a specified orbital inclination this gives the mass ratio of the stars.
\end{itemize}

We have deliberately avoided using a mass-radius relation for properties of the secondary star. Empirical relations exist for the mass donors in CVs (e.g.\ \citealt{SmithDhillon98mn}) but the intrinsic complexity of these systems means that they are based on limited observational data. Theoretical relations for CVs are not trustworthy at present because of the discrepancy between the predicted and observed properties of the CV population (see Sect.~\ref{sec:intro}). Several empirical and theoretical relations exist for late-type main sequence stars which have evolved essentially independently, but recent detailed investigations of detached eclipsing binary systems consistently find that the radii of late-type dwarfs are up to 10\% greater than predicted \citep{TorresRibas02apj,LopezRibas05apj,Ribas05xxx}.

The presence of double-peaked emission suggests that the orbital inclination, $i$, of the SDSS\,J1216 system is quite high. \citet{Warner76iaus} noted that double-peaked emission lines indicate a CV with $i \ga 50^\circ$, although there is no obvious reason why double-peaked lines could not be seen from much lower inclination systems \citep{HorneMarsh86mn}. Choosing an inclination of 60$^\circ$, using our value for the mass function, and assuming a mass of 0.6\Msun\ (1.0\Msun) for the white dwarf, we find that the secondary star has a mass of $0.022 \pm 0.005$\Msun\ ($0.031 \pm 0.005$\Msun). These values are both in reasonable agreement with theoretical predictions of the properties of CVs which have evolved past the orbital period minimum \citep{KolbBaraffe99mn,Howell++01apj}. Adopting a higher inclination would cause the inferred mass of the secondary star to decrease. Finally, using $M_{\rm WD} = 0.69 \pm 0.13$\Msun, as given by \citet{SmithDhillon98mn} for CVs below the period gap, Fig.~\ref{fig:1216mass} indicates that the secondary star of SDSS\,J1216 has a mass in the range 0.012 to 0.040 \Msun\ (including the uncertainty in our measurement of the mass function).

Fig.~\ref{fig:1216mass} shows the effect of the above constraints on the masses of the components of SDSS\,J1216 and the inclination of the system. Light shaded areas indicate those parts of the diagram which are only accessible for an inclination of 75$^\circ$ or more. Dark shaded areas indicate the areas forbidden by the constraint on the maximum Keplerian velocity of the disc. It also shows the constraint due to the mass function for a number of selected inclinations between 20$^\circ$ and 75$^\circ$.

From Fig.~\ref{fig:1216mass} it is clear that there are two possible solutions to the system. The most likely of these, given that the double-peaked lines suggest a high inclination, is a donor star with a mass significantly below the hydrogen-burning limit. The less likely alternative is a white dwarf with a large mass of 1.15\Msun\ or greater. Thus SDSS\,J1216 is a good candidate for a CV containing a brown-dwarf-mass donor star.

SDSS\,J1216 resembles GD\,552, which also has double-peaked lines, with a peak separation smaller than SDSS\,J1216 but a very similar velocity amplitude of $17 \pm 4$\kms\ \citep{HessmanHopp90aa}. Based on their data, \citet{HessmanHopp90aa} suggested that GD\,552 is a low-inclination CV containing a high-mass white dwarf and a main sequence secondary star. In the light of the predictions of theoretical models \citep[e.g.][]{Kolb93aa,KolbBaraffe99mn,Howell++01apj,Politano04apj}, a viable alternative hypothesis is that GD\,552 and SDSS\,J1216 are CVs with substellar donor stars. This type of system is expected to be common and in fact should dominate the galactic CV population. We therefore feel confident in proposing that SDSS\,J1216 is a strong candidate for a CV with a brown-dwarf-mass secondary star.


\section{Discussion and Conclusions}

This work presents phase-resolved spectroscopy of six CVs recently discovered in SDSS spectroscopic survey observations. For four of these systems we have obtained a precise orbital period using radial velocities measured from the wings of H$\beta$ or H$\alpha$ emission lines. Three of the four systems have periods below the period gap: 96\,min for SDSS\,J0233, 82\,min for SDSS\,J1035 and 99\,min for SDSS\,J1216 (although we have not been able to completely rule out the possibility that the 92\,min one-day alias is in fact the correct period). The fourth, SDSS\,J0911, has a period of 296\,min, which is substantially longwards of the period gap.

Whilst we have found a clear 96\,min periodicity in the emission-line radial velocities of SDSS\,J0233, a period of 60\,min is present in the light curves of this system. We suggest that this system is a low-accretion-rate intermediate polar and the 60\,min variation represents the spin period of the white dwarf.

The spectra of SDSS\,J1035 show it to be an eclipsing system; subsequent ULTRACAM photometry has found the eclipses to be total (Littlefair et al., submitted). Its short period and spectrum dominated by a cool white dwarf are reminiscent of WZ\,Sge systems (e.g.\ \citealt{Patterson+05pasp,Templeton06pasp}). From modelling the spectrum we find a white dwarf effective temperature of $11\,700 \pm 1500$\,K and an upper limit of L0 to the spectral type of the secondary star, which is not detected in the spectrum. The high quality of the VLT/FORS2 observations has also allowed Doppler tomograms to be constructed for its accretion disc, and these show a clear bright spot and a disc with comparatively high-velocity motions.

We have found SDSS\,J1216 to have a very low velocity amplitude of $13.8 \pm 1.6$\kms. To our knowledge this is the smallest amplitude so far measured for a CV from emission-line radial velocities (excluding lower limits set by non-detection of velocity variation). Through applying a series of physically reasonable constraints we find that the system must have either a high-mass white dwarf primary component and/or a secondary star with a mass in the brown dwarf regime. Systems with donor masses well below 0.1\Msun\ are expected to be very faint, because the secondary star is dim and the mass transfer rate is low, and so under-represented in the current observational population of CVs.

There are several candidates for CVs with brown-dwarf-mass secondary stars and their properties have been reviewed by \citet{Littlefair++03mn}, to whose list should be added the recently-studied CVs HS\,2331+3905 \citep{Araujo+05aa} and RE\,J1255+266 \citep{Patterson++05pasp}. In general, inferences of the secondary mass have been drawn from the spectral energy distribution of the system, from calibrations between superhump period excess and mass ratio, or through detection of spectral lines of the secondary component in a few cases. \citet{Littlefair++03mn} found that direct evidence for a brown dwarf secondary does not exist for any CV, but that some systems have substantial indirect evidence.

Three of the four CVs studied here have very short orbital periods, which is consistent with the expectation that studying successively fainter samples of CVs means that the targets become on average intrinsically fainter and of shorter period \citep{Patterson++05pasp,Gansicke05aspc}. Our ongoing research program of CVs discovered by the SDSS will give orbital periods of a larger homogeneous sample of these systems, allowing us to accurately assess the biases affecting the observed sample of CVs and so accurately measure the properties of the intrinsic population of these objects.

%
%
\section*{Acknowledgements}

The reduced spectra and radial velocity observations presented in this work will be made available at the CDS ({\tt http://cdsweb.u-strasbg.fr/}) and at {\tt http://www.astro.keele.ac.uk/$\sim$jkt/}.

This work is based on observations obtained at the European Southern Observatory Very Large Telescope and FORS2 spectrograph, resulting from ESO proposal 076.D-0326A.1. It is also based in part on observations made with the William Herschel Telescope and the Isaac Newton Telescope, which are operated on the island of La Palma by the Isaac Newton Group in the Spanish Observatorio del Roque de los Muchachos (ORM) of the Instituto de Astrof\'{\i}sica de Canarias (IAC) and on observations made with the Liverpool Telescope which is operated on the island of La Palma by Liverpool John Moores University at the ORM of the IAC. The WHT, INT, and LT data were obtained as part of the 2004 International Time Programme of the night-time telescopes at the European Northern Observatory.

We would like to thank the referee, John Thorstensen, for a very timely and helpful report. JS and SPL acknowledge financial support from PPARC in the form of postdoctoral research assistant positions. BTG acknowledges financial support from PPARC in the form of an advanced fellowship. TRM was supported by a PPARC senior fellowship during the ourse of this work. DDM acknowledges financial support from the Italian Ministry of University and Research (MIUR). PS acknowledges support from NSF grant AST 02-05875.

The following internet-based resources were used in research for this paper: the ESO Digitized Sky Survey; the NASA Astrophysics Data System; the SIMBAD database operated at CDS, Strasbourg, France; and the ar$\chi$iv scientific paper preprint service operated by Cornell University.

Funding for the Sloan Digital Sky Survey (SDSS) has been provided by the Alfred P.\ Sloan Foundation, the Participating Institutions, the National Aeronautics and Space Administration, the National Science Foundation, the U.\ S.\ Department of Energy, the Japanese Monbukagakusho, and the Max Planck Society. The SDSS website is http://www.sdss.org/.


\bibliographystyle{mn_new}
\bibliography{aamnem99,jkt}
\bsp
\label{lastpage}

\end{document}